\documentclass[aps,showpacs]{revtex4}

\usepackage{latexsym}
\usepackage{amsmath}
\usepackage{epsfig}

\def\vec#1{\boldsymbol{#1}}

\begin{document}
\title{Z-scaling and space-time structural relativity}

\author{I. Zborovsk\'{y}}
\email{zborovsky@ujf.cas.cz}
\affiliation{Nuclear Physics Institute, Academy of Sciences of the Czech Republic,
\v{R}e\v{z}, Czech Republic}

\begin{abstract}
Assuming fractality of hadronic constituents,
we introduce elements of special realization of the relativity principle applied
to physical quantities expressed with respect to various fractal structures.
The construction is inspired by the premisses of the z-scaling observed in the
inclusive reactions at high energies.
The scheme concerns parton structure of hadrons and nuclei at small scales.
\end{abstract}


\maketitle

\section{Introduction}

The inclusive spectra of secondaries produced with large transverse momenta
in high energy collisions of hadrons and nuclei
provide unique information about the properties of quark and gluon interactions.
As follows from numerous studies in relativistic physics, common feature of these
processes is local character of the hadron interactions.
This leads to a conclusion about
dimensionless constituents participating in the collisions.
Fact that the interaction is local manifests naturally
in a scale-invariance of the interaction cross sections.
The invariance is a special case of the self-similarity
\cite{Matveev,Brodsky} which enables to predict and study
various phenomenological regularities reflecting the point-like
nature of the underlying interactions.

The particle spectra  are often presented \cite{McLerran}
as a scaling function depending on the transverse mass $m_{\bot}=\sqrt{p^2_{\bot}+m^2}$.
Such regularity in the behaviour of the differential cross
sections concerns the central interaction region and holds in the limited
range of the transverse momenta ($p_{\bot}\lesssim 3$~GeV/c).
For larger momenta ($p_{\bot}\lesssim 6$~GeV), the scaling can be preserved
in the variable $m_{\bot}/K$,  when introducing an energy dependent scale $K(s)$ \cite{Hwa}.
In analogy with the Koba-Nielsen-Olesen scaling \cite{KNO} of the multiplicity
distributions, the energy dependence of the scale $K(s)$ for the inclusive reactions was
identified \cite{Z96} with the energy dependence of the average multiplicity density
$dN(0)/d\eta$ of particles produced in the central region of the interaction.
The concept of self-similarity of hadron interactions at constituent level was complemented
by considerations about fractal character \cite{Mandelbrot} of the objects undergoing the
high energy collisions.
This lead to introduction of the scaling variable \cite{Z99}
\begin{equation}
z = z_0\Omega^{-1},
\label{eq:x1}
\end{equation}
where
\begin{equation}
\Omega(x_1,x_2) = (1-x_1)^{\delta_1}(1-x_2)^{\delta_2}.
\label{eq:x2}
\end{equation}
The variable $z$ has character of a fractal measure.
For a given production process,
its finite part $z_0$ is ratio of the transverse energy released in the underlying collision
of constituents and the average multiplicity density $dN(0)/d\eta$.
The divergent part $\Omega^{-1}$
describes resolution at which the  collision of
the constituents can be singled out of this process.
The $\Omega(x_1,x_2)$ represents relative number of all initial
configurations containing the constituents which carry the
fractions $x_1$ and $x_2$ of the incoming momenta.
The $\delta_1$ and $\delta_2$ are anomalous fractal dimensions of the colliding
objects (hadrons and nuclei).
Divergent character of the variable $z$ secures that there is no $z_{max}$ limit for
any energy where the scaling has to be a'priori violated.
The scaling function was  found to be independent of the center-of-mass energy
and the angle of produced particles over a vide kinematic range \cite{Z99}.
The energy and angular independence of the $z$ scaling was shown for the production
of high $p_{\bot}$ jets \cite{Z00}.
A-universality of the scaling was demonstrated for $pA$ collisions for
various nuclei \cite{Z01}.

The goal of the paper is to focus on general premisses
of the $z$ scaling which concern  fractality of the constituent interactions.
Fractals are mathematical concepts expressing the self-similarity and inexhaustible
structure at small scales \cite{Mandelbrot}.
The geometrical objects model the internal parton structure of
hadrons and nuclei revealed in their interactions still more and more
with increasing energies.
This property encountered in modern physics is connected with
scale dependence of physical laws gradually
emerging in various experimental and theoretical investigations \cite{Wheeler,Hawking}.
Such extension of physics is intrinsically linked to the
evolution of the concept of space-time
\cite{Ord,Nottale,Pissondes,Ellis}.
Its structure is characterized by explicitly
scale dependent metric potentials. Asking questions about
the metrics leads one to question the relativity.
The relativistic principle besides motion, applies also to the laws of scale \cite{Nottale}.
The scale changes are expressed in terms of the "scale velocity" defined in a special way.
In this paper we attempt to extend the realization of the relativity
principle to various space-time structures emerging at small scales.
Change of the structures is characterized by a "structural velocity"
expressed in ratios of their anomalous fractal dimensions.
Similarly as the scale velocity, the structural velocity does not represent
any real motion. While first characterizes change of the state of scale, later
expresses change of the structures at a given scale.

The paper is organized as follows.
The description of the parton interactions as used by
the construction of the $z$ scaling is recapitulated in Sec. II.
Elements of the structural relativity for isolated fractal reference systems
are introduced in Sec. III.
In Sec. IV., we consider structural anisotropy of space-time induced by the interaction
of the fractal objects possessing mutually different anomalous fractal dimensions.
Relativistic mechanics in anisotropic space-time is discussed in Sec. V.
Relations between the variables used in the kinematical and mechanical sector
are presented in Sec. VI. Klein-Gordon equation and the conclusions are
discussed in Sec. VII., VIII, and in the Appendix.

\vskip 0.5cm
{\section {Constituent interactions}}

At sufficient high energies, the interactions of hadrons and nuclei
can be considered as an ensemble of individual interactions
of their constituents.
The constituents are partons in the parton model or
quarks and gluons which are the building blocks in the theory of QCD.
Production of particles with large transverse momenta
from such reactions has relevance
to physics at small interaction distances.
In this region, the interactions of hadronic constituents
are local relative to the resolution which depends on the
kinematical characteristics of particles produced in the
collisions.
In accordance with the property of locality it has been suggested
\cite{Stavinsky} that gross features of the
single-inclusive particle distributions for the reaction
\begin{equation}
m_A+m_B \rightarrow m_c + X
\label{eq:x3}
\end{equation}
can be described in terms of the corresponding kinematical
characteristics of the sub-process
\begin{equation}
(x_1m_A) + (x_2m_B) \rightarrow m_c +
(x_1m_A+x_2m_B + \bar{m}_c)
\label{eq:x4}
\end{equation}
which is subject to the condition
\begin{equation}
(x_1p_A + x_2p_B - p_c)^{2} = (x_1m_A + x_2m_B + \bar{m}_c)^{2} .
\label{eq:x5}
\end{equation}

The $x_{1}$ and $x_{2}$ are fractions of the
incoming four-momenta $p_A$ and $p_B$ of the colliding objects with the masses
$m_A$ and $m_B$.
The $p_c$  is four-momentum of the inclusive particle
with the mass $m_c$.
The parameter $\bar{m}_c$ is minimal mass used in connection with the
internal conservation laws (for isospin, baryon number, and strangeness).
The relationship (\ref{eq:x5}) can be conveniently written in the form
\begin{equation}
x_1x_2-x_1\lambda_2-x_2\lambda_1-\lambda_0 = 0,
\label{eq:x6}
\end{equation}
where
\begin{equation}
\lambda_1 = \frac{(p_Bp_c)+m_B\bar{m}_c}{(p_Ap_B)-m_Am_B} ,\ \ \ \ \
\lambda_2 = \frac{(p_Ap_c)+m_A\bar{m}_c}{(p_Ap_B)-m_Am_B} ,\ \ \ \ \
\lambda_0 = \frac{0.5(\bar{m}_c^2-m_c^2)}{(p_Ap_B)-m_Am_B} .
\label{eq:x7}
\end{equation}

We have determined \cite{Z99} the momentum fractions $x_1$ and $x_2$ in the way to
minimize the resolution $\Omega^{-1}$ of the fractal measure $z$ with respect to all possible
sub-processes (\ref{eq:x4}) which can lead to production of the inclusive particle with
the four-momentum $p_c$.
This corresponds to maximum of the functional
\begin{equation}
F(x_1,x_2) = \Omega(x_1,x_2) +
\beta(x_1x_2-x_1\lambda_2-x_2\lambda_1-\lambda_0)
\label{eq:x8}
\end{equation}
with a Lagrange multiplicator $\beta$.
The momentum fractions resulting
from this requirement have the form
\begin{equation}
x_1 = \lambda_1 + \chi_1 , \ \ \ \ \ \ \
x_2 = \lambda_2 + \chi_2
\label{eq:x9}
\end{equation}
where
\begin{equation}
\chi_1 = \sqrt{\mu_1^{2}+\omega_1^{2}}-\omega_1 , \ \ \ \ \ \
\chi_2 = \sqrt{\mu_2^{2}+\omega_2^{2}}+\omega_2 .
\label{eq:x10}
\end{equation}
Here we have used the notations
\begin{equation}
\mu_1^2 = \lambda^2\alpha\frac{(1-\lambda_1)}{(1-\lambda_2)} , \ \ \ \ \ \
\mu_2^2 = \lambda^2\frac{1}{\alpha}\frac{(1-\lambda_2)}{(1-\lambda_1)},
\label{eq:x11}
\end{equation}
\begin{equation}
\omega_1 = \lambda\frac{(\alpha-1)}{2(1-\lambda_2)} , \ \ \ \ \ \
\omega_2 = \lambda\frac{(\alpha-1)}{2\alpha(1-\lambda_1)},
\label{eq:x12}
\end{equation}
\begin{equation}
\lambda = \sqrt{\lambda_1\lambda_2+\lambda_0}.
\label{eq:x13}
\end{equation}
The structural parameter $\alpha=\delta_2/\delta_1$ is ratio of the anomalous fractal
dimensions of the colliding objects.
Procedure of minimizing the fractal resolution
$\Omega^{-1}$ leads to the following consequences.
The non-trivial result is that $\omega_i$ and $\mu_i$ are both
related by the formulae
\begin{equation}
\omega_1 = \mu_1U , \ \ \ \ \ \ \ \ \ \
\omega_2 = \mu_2U
\label{eq:x14}
\end{equation}
through the same value
\begin{equation}
U = \frac{\alpha-1}{2\sqrt{\alpha}}\xi.
\label{eq:x15}
\end{equation}
The quantity $U$ consists of  $\alpha$ dependent structural part and a kinematical
factor
\begin{equation}
\xi = \frac{\lambda}{ \sqrt{ (1-\lambda_1)(1-\lambda_2)}} .
\label{eq:x16}
\end{equation}
The factor $\xi\le 1$ is function of the center-of-mass
energy and momenta of the observed secondaries.
For the inclusive reactions, it characterizes the scale resolution.
When approaching the phase-space
limit of the reaction (\ref{eq:x3}), the $\xi$ tends to unity.
Along the whole phase-space limit $x_1=x_2=1$ and $\xi=1$.
The phase-space boundary corresponds to the fractal limit
with the infinite resolution $\Omega^{-1}$. The fractal limit is thus equivalent
to infinite value of the fractal measure $z$. This extreme reflects situation when
the whole reaction (\ref{eq:x3}) degenerates to the single sub-process (\ref{eq:x4}).
Though kinematically accessible at any centre-of-mass energy, its probability
is null. What is of physical meaning is the way the probability approaches this limit.
The z-presentation of experimental data reveals power
dependence of the scaling behaviour \cite{Z99,Z00,Z01} in the range of large $z$
suggesting specific values of the anomalous fractal dimensions $\delta_i$.

The expressions (\ref{eq:x10}) and (\ref{eq:x14}) imply $\chi_1\chi_2=\mu_1\mu_2$.
Moreover, while  $\chi_i$ and $\mu_i$ obtained by the minimalization procedure
of $\Omega^{-1}$ are non-trivial functions of structural parameter $\alpha$,
the combination
\begin{equation}
\chi_1\chi_2 = \mu_1\mu_2 = \lambda^2
\label{eq:x17}
\end{equation}
does not depend on $\alpha$.
This allows to write the sub-process (\ref{eq:x4}) in the symbolic form
\begin{equation}
(\lambda_1+\chi_1) + (\lambda_2+\chi_2) \rightarrow
(\lambda_1+\lambda_2)+(\chi_1+\chi_2).
\label{eq:x18}
\end{equation}
Equation (\ref{eq:x17}) reflects the transverse momentum balance of this sub-process.
The last relation should be understood that $\lambda_i$ parts of the interacting
constituents underly the production of the inclusive particle, while the
$\chi_i$ parts are responsible for the creation of its recoil.
Other formal consequences resulting from minimal resolution in the fractal measure
$z$ will be discussed bellow.

\vskip 0.5cm
{\section {Space-time structural relativity}}

This section is devoted to the essential points to be clarified in our
approach. It is based on suggestion that besides the kinematical
variables there exist structural degrees of freedom to be taken into account
for description of the hadronic constituent sub-processes at small scales. The
construction relies on the constituent fractal-like compositeness as
an universal property of hadronic matter.
In this view, we focus on specific properties of the kinematical and
mechanical variables defined relative to coordinate systems connected with
different fractal structures.
It will be shown that transformations between these variables form a group
with the corresponding composition rules.
The properties will be studied with respect to the "structural velocity"
defined sa follows
\begin{equation}
\vec{u} =\frac{\vec{U}}{\sqrt{1+U^2}}
\label{eq:x19}
\end{equation}
where $\vec{U}=(0,0,U)$ is given by Eq. (\ref{eq:x15}). Here we have oriented
third coordinate axis in the direction of the collision beams.
The inverse relation reads
\begin{equation}
\vec{U} =\frac{\vec{u}}{\sqrt{1-u^2}}.
\label{eq:x20}
\end{equation}
This enables us to consider the quantities $U_i$ as space components of structural
four-velocity.
The structural velocity $\vec{u}$ has its origin in the structural asymmetry of
the interaction and vanishes in the collision of the fractal objects possessing  equal
anomalous fractal dimensions $\delta_1=\delta_2$.
The quantity does not represent real motion but
characterizes structural polarization in the interaction region.
As the formula (\ref{eq:x15}) for $U$ is Lorenz invariant with respect to motion,
the value of the structural velocity is the same when evaluated in whatever
inertial reference frame.
This applies equally to any expressions which otherwise can depend on
$\vec{u}$ but are Lorenz invariants relative to motion. Such expressions
can be evaluated in arbitrary motion inertial reference frame using the standard methods.

\vskip 0.5cm
{\subsection {Structural relativistic transformations in 1+1 dimensions}}

Exploiting the definition (\ref{eq:x19}),
we can rewrite the expressions (\ref{eq:x10}) as follows
\begin{eqnarray}
\mu_1\!-\!\mu_2 &=& \frac{1}{\sqrt{1\!-\!u^2}}
\left[(\chi_1\!-\!\chi_2) + u (\chi_1\!+\!\chi_2)\right],
\nonumber \\
\mu_1\!+\!\mu_2 &=& \frac{1}{\sqrt{1\!-\!u^2}}
\left[(\chi_1\!+\!\chi_2) + u (\chi_1\!-\!\chi_2)\right],
\label{eq:x21}
\end{eqnarray}
or equivalently
\begin{eqnarray}
\chi_1\!-\!\chi_2 &=& \frac{1}{\sqrt{1\!-\!u^2}}
\left[(\mu_1\!-\!\mu_2) - u (\mu_1\!+\!\mu_2)\right],
\nonumber \\
\chi_1\!+\!\chi_2 &=& \frac{1}{\sqrt{1\!-\!u^2}}
\left[(\mu_1\!+\!\mu_2) - u (\mu_1\!-\!\mu_2)\right].
\label{eq:x22}
\end{eqnarray}
These relations have form of Lorenz transformations along the third coordinate axis
with respect to the structural velocity $u$. According to Eq. (\ref{eq:x17}),
their invariant
\begin{equation}
(\chi_1\!+\!\chi_2)^2-(\chi_1\!-\!\chi_2)^2 =
(\mu_1\!+\!\mu_2)^2-(\mu_1\!-\!\mu_2)^2
\label{eq:x23}
\end{equation}
does not depend on the structural parameter $\alpha=\delta_2/\delta_1$.
Therefore, for a given kinematical factor $\xi$, Eqs. (\ref{eq:x21}) and (\ref{eq:x22})
can be considered as relativistic transformations
of $(\chi_1-\chi_2,\chi_1+\chi_2)$ and $(\mu_1-\mu_2,\mu_1+\mu_2)$ expressed
relative to the reference systems connected with fractal structures of different
anomalous dimensions.
As the combinations of fractions are Lorenz invariants with respect to motion,
we can evaluate both of them in the center-of-mass system of the reaction (\ref{eq:x3}).
This gives
\begin{equation}
\mu_1+\mu_2 = \frac{2}{\sqrt{s}}E_1, \ \ \ \
\mu_1-\mu_2 = \frac{2}{\sqrt{s}}p_{1z}, \ \ \ \
\sqrt{\mu_1\mu_2} =\frac{1}{\sqrt{s}}\sqrt{p^2_{1\bot}+\bar{m}_c^2}
\label{eq:x24}
\end{equation}
or
\begin{equation}
\chi_1+\chi_2 = \frac{2}{\sqrt{s}}E_2, \ \ \ \
\chi_1-\chi_2 = \frac{2}{\sqrt{s}}p_{2z}, \ \ \ \
\sqrt{\chi_1\chi_2} =\frac{1}{\sqrt{s}}\sqrt{p^2_{2\bot}+\bar{m}_c^2} .
\label{eq:x25}
\end{equation}
Here we have denoted $E_1$, $\vec{p}_1$ and $E_2$, $\vec{p}_2$ the center-of-mass
energy and momentum of the recoil object $\bar{m}_c$ expressed relative to the
structural reference frames $S_1$ and $S_2$, respectively.
The single structural frames are associated with isolated fractal structures
characterized by the  anomalous fractal dimensions $\delta_1$ and $\delta_2$.
The reference system $S_1$ is connected with the fractal structure of the first
fractal object (nucleus A) and the reference system $S_2$ with the second
fractal object (nucleus B).
If both fractal structures possess mutual different anomalous dimensions
$(\delta_1\ne\delta_2)$,
the variables are linked by the relativistic transformations
\begin{eqnarray}
p_{2z} &=& \frac{1}{\sqrt{1\!-\!u^2}}\left(p_{1z} - u E_1\right),
\nonumber \\
E_2 &=& \frac{1}{\sqrt{1\!-\!u^2}}\left(E_1 - u p_{1z}\right)
\label{eq:x26}
\end{eqnarray}
depending on the structural velocity $u\ne 0$.
The invariant of these transformations, $E^2_2-p^2_{2z} = E^2_1-p^2_{1z}$,
corresponds to the invariant form (\ref{eq:x23}).

Let us now examine fractal limit which is equivalent to the infinite resolution.
As follows from the definition of the scaling variable $z$, the fractal limit
can be achieved kinematically at any energy. It corresponds to the
phase-space limit with $x_1=x_2=1$.
In this case, the fractions $\chi_i$ approach their limiting values
$\chi_1=\cos^2(\theta_B/2)$ and $\chi_2=\sin^2(\theta_B/2)$ where
$\theta_B$ is the center-of-mass angle of the recoil particle
expressed relative to the structural reference frame $S_2$.
The corresponding kinematics is given by the sphere
\begin{equation}
\chi_1-\chi_2=\cos\theta_B ,
\ \ \ \ \ \ \ \
\chi_1+\chi_2=1
\label{eq:x27}
\end{equation}
in this frame. When transforming the sphere to the reference system associated with the
fractal structure of the object $A$, we get
\begin{eqnarray}
\mu_1-\mu_2 &=&
\sqrt{\alpha}\sin^2(\theta_B/2)-
\frac{1}{\sqrt{\alpha}}\cos^2(\theta_B/2),
\nonumber \\
\mu_1+\mu_2 &=&
\sqrt{\alpha}\sin^2(\theta_B/2)+
\frac{1}{\sqrt{\alpha}}\cos^2(\theta_B/2) .
\label{eq:x28}
\end{eqnarray}
These equations represent angular parameterization of an
ellipse with the focus in the origin of the reference system.
Similarly, the spherical kinematics
\begin{equation}
\mu_1-\mu_2=\cos\theta_A ,
\ \ \ \ \ \ \ \ \ \
\mu_1+\mu_2=1
\label{eq:x29}
\end{equation}
expressed in terms of the center-of-mass angle $\theta_A$ of the recoil particle $\bar{m}_c$
in the reference frame $S_1$ transforms to the ellipse
\begin{eqnarray}
\chi_1-\chi_2 &=&
\frac{1}{\sqrt{\alpha}}\sin^2(\theta_A/2)-
\sqrt{\alpha}\cos^2(\theta_A/2),
\nonumber \\
\chi_1+\chi_2 &=&
\frac{1}{\sqrt{\alpha}}\sin^2(\theta_A/2)+
\sqrt{\alpha}\cos^2(\theta_A/2)
\label{eq:x30}
\end{eqnarray}
in the fractal reference system $S_2$.
The transverse components $\chi_{\bot} = \sin\theta_B= \sin\theta_A =\mu_{\bot}$
are conserved by these structural transformations.

To end up with infinite resolution, we mention one remarkable property
concerning the composition of the structural velocities in 1+1 dimensions.
As the kinematical scale factor (\ref{eq:x16})
is unity in the fractal limit ($\xi=1$), the
relation (\ref{eq:x15}) takes the simple form
\begin{equation}
U = \frac{\alpha-1}{2\sqrt{\alpha}} = \frac{u}{\sqrt{1-u^2}}.
\label{eq:x31}
\end{equation}
When solving this equation with respect to $u$, we get the structural velocity
\begin{equation}
u = \frac{\alpha-1}{\alpha+1}
\label{eq:x32}
\end{equation}
as an exclusive function of the ratio $\alpha=\delta_2/\delta_1$ of the
anomalous fractal dimensions of the colliding objects.
This relation satisfies the standard relativistic composition rule
\begin{equation}
u_a =
\frac{u+u_b}{1+uu_b} ,
\label{eq:x33}
\end{equation}
provided
\begin{equation}
\alpha_a = \alpha\alpha_b.
\label{eq:x34}
\end{equation}
The conclusion one has to make for the fractal limit is following.
While the composition of structural
velocities in 1+1 dimensions is governed by Einstein-Lorenz law, the composition of the
corresponding ratio of the anomalous fractal dimensions obeys
the multiplicative group law.
Such correspondence is specific expression of
structural relativity in which single fractal structures play
analogous roles as the inertial systems in the motion relativity.

\vskip 0.5cm
{\subsection {Structural relativistic transformations in 3+1 dimensions}}

In this section we discuss generalization of the structural relativistic
transformations (\ref{eq:x26}) for 3 spatial dimensions.
First we consider the transformations which left untouched the variables
transverse to the relativistic boost.
This is characteristic for the Lorenz transformations
\begin{eqnarray}
\vec{p}_2 &=& \vec{p}_1
+\vec{u}\left(\frac{\vec{u}\vec{p}_1}{u^2}(\gamma-1)-\gamma E_1\right),
\nonumber \\
E_2 &=& \gamma\left(E_1 - \vec{u}\vec{p}_1\right) .
\label{eq:x35}
\end{eqnarray}
As the transformations concern the structural relativity, the relativistic factor
\begin{equation}
\gamma=(1-u^2)^{-1/2}
\label{eq:x36}
\end{equation}
is given in terms of the structural velocity $\vec{u}$.
In order to preserve the standard relativistic relations in both fractal reference
frames $S_1$ and $S_2$, we have to require the same transformations also for
coordinates and time,
\begin{eqnarray}
\vec{r}_2 &=& \vec{r}_1
+\vec{u}\left(\frac{\vec{u}\vec{r}_1}{u^2}(\gamma-1)-\gamma t_1\right),
\nonumber \\
t_2 &=& \gamma\left(t_1 - \vec{u}\vec{r}_1\right) .
\label{eq:x37}
\end{eqnarray}
Last equations entail composition of the structural velocity
$\vec{u}$ with the motion velocity $\vec{v}_1=d\vec{r}_1/dt_1$ in formally
standard way \cite{Moller}
\begin{equation}
\vec{v}_2 = \frac{\gamma^{-1}\vec{v}_1
+\vec{u}\left[
\left(1\!-\!\gamma^{-1}\right)
\vec{u}\!\cdot\!\vec{v}_1/u^2
-1\right]}
{1-\vec{u}\!\cdot\!\vec{v}_1}.
\label{eq:x38}
\end{equation}
The result is the motion velocity $\vec{v}_2=d\vec{r}_2/dt_2$ with respect to the
fractal reference system $S_2$.
The composition of the structural velocities alone is given by the group
structure of the structural transformations (\ref{eq:x37}).
When using the four-dimensional notation $r_2=\Lambda(\vec{u})r_1$
($p_2=\Lambda(\vec{u})p_1$) with
\begin{equation}
\Lambda (\vec{u}) =
\left(
\begin{array}{cc}
\delta_{ij}\!+(\gamma-1) u_iu_j/u^{2} &
-\gamma u_i  \\
-\gamma u_j  &  \gamma\\
\end{array}
\right)  ,
\label{eq:x39}
\end{equation}
the group structure of the Lorenz transformations is expressed as follows
\begin{equation}
R(\vec{\phi})\Lambda(\vec{u}_a) =
\Lambda(\vec{u}_b) \Lambda(\vec{u}) .
\label{eq:x40}
\end{equation}
The matrix
\begin{equation}
R(\vec{\phi}) =
\left(
\begin{array}{cc}
r_{ij} & 0 \\
0 & 1  \\
\end{array}
\right)
\label{eq:x41}
\end{equation}
describes three dimensional Thomas precession \cite{Thomas} around the vector
$\vec{\phi}\sim \vec{u}_b\!\times\!\vec{u}$ known in the theory of relativity.
The corresponding composition of the structural velocities
defined with respect to various isolated fractal reference frames reads
\begin{equation}
\vec{u}_b = \frac{\gamma^{-1}\vec{u}_a
+\vec{u}\left[
\left(1\!-\!\gamma^{-1}\right)
\vec{u}\!\cdot\!\vec{u}_a/u^2
-1\right]}
{1-\vec{u}\!\cdot\!\vec{u}_a}.
\label{eq:x42}
\end{equation}
Though formally identical, this should be distinguished from the composition of the
motion velocities,
$\vec{v}''_i=\vec{v}'_i\ominus\vec{v}_i,$ following from the relations
$r''_i=\Lambda(\vec{v}_i)r'_i, i=1,2$.
For the situation considered above we conclude that the structural velocities
and the motion velocities are composed separately and mutually in the same
way.

There exists another generalization of Eqs. (\ref{eq:x26}) connected
with mutual spinning of the interacting fractal structures around the collision axis.
This transformation has the form
\begin{eqnarray}
\vec{p}_2 &=& \gamma^{-1}\vec{p}_1-\sigma\vec{u}\!\times\!\vec{p}_1-
\gamma\vec{u}(E_1-\vec{u}\!\cdot\!\vec{p}_1),
\nonumber \\
E_2 &=& \gamma(E_1-\vec{u}\!\cdot\!\vec{p}_1)
\label{eq:x43}
\end{eqnarray}
where $\sigma=\pm 1$ corresponds to the right/left spinning (or torsion) of the
colliding fractals.
For similar reasons as above, the same should apply to the kinematical
variables $\vec{r}$ and $t$,
\begin{eqnarray}
\vec{r}_2 &=& \gamma^{-1}\vec{r}_1-\sigma\vec{u}\!\times\!\vec{r}_1-
\gamma\vec{u}(t_1-\vec{u}\!\cdot\!\vec{r}_1),
\nonumber \\
t_2 &=& \gamma(t_1-\vec{u}\!\cdot\!\vec{r}_1).
\label{eq:x44}
\end{eqnarray}
The inverse transformations are obtained by the interchange $1\leftrightarrow 2$
with replacing $\vec{u}$ by $-\vec{u}$.
The above transformations preserve the invariant forms $E^2-p^2 =m_0^2$
and $t^2-r^2=\tau^2$. The relations
\begin{equation}
E_1=\frac{m_0}{\sqrt{1-v_1^2}}, \ \ \ \ \
\vec{p}_1 = E_1\vec{v}_1, \ \ \ \ \ \ \ \
E_2=\frac{m_0}{\sqrt{1-v_2^2}}, \ \ \ \ \
\vec{p}_2 = E_2\vec{v}_2
\label{eq:x45}
\end{equation}
are thus automatically fulfilled.
This secures that the standard motion relativity remains valid in both fractal
reference frames $S_1$ and $S_2$.
The relativistic transformations with respect to motion, $r''_i=\Lambda(\vec{v}_i)r'_i$
$(p''_i=\Lambda(\vec{v}_i)p'_i)$, i=1,2, imply the usual Lorenz composition of the
motion velocities, $\vec{v}''_i=\vec{v}'_i\ominus\vec{v}_i,$ also in this case.
However, according to Eq. (\ref{eq:x44}), the motion velocities $\vec{v}_1$ and $\vec{v}_2$
expressed relative to the structural reference systems $S_1$ and $S_2$
are linked by the expressions
\begin{equation}
\vec{v}_2 = \frac{\vec{v}_1 -\sigma\gamma\vec{u}\!\times\!\vec{v}_1}
{\gamma^2(1-\vec{u}\!\cdot\!\vec{v}_1)} - \vec{u}, \ \ \ \ \ \ \
\vec{v}_1 = \frac{\vec{v}_2 +\sigma\gamma\vec{u}\!\times\!\vec{v}_2}
{\gamma^2(1+\vec{u}\!\cdot\!\vec{v}_2)} + \vec{u}
\label{eq:x46}
\end{equation}
which differ from Eq. (\ref{eq:x38}). Similarly, the composition of the structural
velocities alone is different here. It is given by the group structure of the
structural transformations (\ref{eq:x43}) and (\ref{eq:x44}).
When using the four-dimensional notation $r_2=\Pi(\vec{u})r_1$
($p_2=\Pi(\vec{u})p_1$) with
\begin{equation}
\Pi (\vec{u}) =
\left(
\begin{array}{cc}
\gamma^{-1}\delta_{ij}\!+\sigma\epsilon_{ijk}u_k + \gamma u_iu_j &
-\gamma u_i  \\
-\gamma u_j  &  \gamma\\
\end{array}
\right) ,
\label{eq:x47}
\end{equation}
the group structure of these transformations is expressed by the relation
\begin{equation}
R(\vec{\psi})\Pi(\vec{u}_a) =
\Pi(\vec{u}_b) \Pi(\vec{u}) .
\label{eq:x48}
\end{equation}
The corresponding structural velocities are composed as follows
\begin{equation}
\vec{u}_b = \frac{\vec{u}_a -\sigma\gamma\vec{u}\!\times\!\vec{u}_a}
{\gamma^2(1-\vec{u}\!\cdot\!\vec{u}_a)} - \vec{u}.
\label{eq:x49}
\end{equation}
Really, one can compute the matrix product
$\Pi(\vec{u}_b) \Pi(\vec{u})\Pi(-\vec{u}_a)$ and find that the result
has the structure (\ref{eq:x41}). One can convince itself also, that the
corresponding spatial part $r_{ij}$ has the form
\begin{equation}
r (\vec{\psi}) =
\left(
\begin{array}{ccc}
n_1^2+(1\!-\!n_1^2)\cos\psi &
n_1n_2(1\!-\!\cos\psi)-n_3\sin\psi &
n_1n_3(1\!-\!\cos\psi)+n_2\sin\psi  \\
n_1n_2(1\!-\!\cos\psi)+n_3\sin\psi &
n_2^2+(1\!-\!n_2^2)\cos\psi &
n_2n_3(1\!-\!\cos\psi)-n_1\sin\psi  \\
n_1n_3(1\!-\!\cos\psi)-n_2\sin\psi &
n_2n_3(1\!-\!\cos\psi)+n_1\sin\psi &
n_3^2+(1\!-\!n_3^2)\cos\psi \\
\end{array}
\right)
\label{eq:x50}
\end{equation}
representing general parametrization of 3D rotation around an unit
vector $\vec{n}=\vec{\psi}/\psi$.
Unlike the standard Thomas precession, the precession (\ref{eq:x50})
is non-zero even for composition of the collinear velocities.
This circumstance is connected with existence of the
vector product in the transformation equations.
Once convincing ourselves in the above expressions,
we can state that the structural transformations (\ref{eq:x43}) and (\ref{eq:x44}) form a group.
The corresponding composition of the structural velocities includes term characterising
torsion.
While the motion velocities alone are composed in the standard Lorenzian way,
the composition including at lest one structural velocity results in
formula which contains the term with torsion.

The transformations (\ref{eq:x35}) and (\ref{eq:x37}), or (\ref{eq:x43}) and (\ref{eq:x44})
represent the mathematical expression of special realization  of the space-time
structural relativity. They concern the relativity with respect to the
self-similar scale structures which are modeled by fractals of various
anomalous dimensions. The corresponding structural transformations relate
physical quantities given in one fractal reference frame with the quantities
expressed relative to the other one. Single isolated reference frames associated with
fractal structures of different anomalous fractal dimensions play analogous role
as the inertial systems in the motion relativity.
The above relativistic transformations suggest
that there does not exist any absolute structural reference system
connected either with a fractal object or with a particular structure of the (QCD) vacuum.

\vskip 0.5cm
{\section {Induced anisotropy of space-time}}

In our construction we associate single fractal reference systems with
the extended structural objects colliding at high energies. Their fractal structure
models parton content of hadrons and nuclei at small scales. It concerns
subtle net of quarks, anti-quarks and gluons revealed still more and more with increasing
resolution.
Pursuing the ideas of space-time structural relativity, we discuss consequences
in the proposed thinking frame.
The consequences lay beyond the relativistic transformations
(\ref{eq:x35}) and (\ref{eq:x37}), or (\ref{eq:x43}) and (\ref{eq:x44})
which are the transformations connecting two isolated structural systems.
Beside the fractal objects, one has to consider general frames related
to any structure of space-time as well.
According to the accepted notions of quantum field
theories, the space-time vacuum is not an empty space.
Its intimate structure is governed by the same processes
which influence the very structure of hadrons and nuclei. The
self-similarity and infinity of the elementary creation and annihilation
processes allows us to consider the vacuum in the
framework of fractal geometry as well.
The way through which space-time properties are related to matter properties is
instructive. It consists in attributing to space-time those properties of
matter which are universal. It was suggested by many authors
(see e.g. \cite{Ord,Nottale,Pissondes}),
that one of such universal property is fractality,
the never ending self-similar content of matter forming its intimate structure at small scales.
In this view we go beyond the isolated fractal objects and generalize our working hypothesis
as follows: Ones we adopt fractality of hadronic constituents and consider ultra-relativistic
collisions of hadrons and nuclei as collisions of fractals, we can conjecture
that interaction of these fractal objects induces deformation to the very structure of
space-time. If the colliding objects possess mutually different
anomalous fractal dimensions ($\delta_1\ne \delta_2$), it is natural to imagine
that, due to fractality,  vacuum structure acquires a polarization (or anisotropy)
along the collision axis.
Anisotropy of space-time induced by the interaction is visualized  as a fractal
background or sort of a fractal medium "moving" with the structural
velocity $\vec{u}$.
The elementary constituent interactions take place on this background in
disturbed space-time. As far as our working hypothesis.

One of the attributes of scale dependent fractal space-time is the fundamental
consequence, namely breaking of the reflection invariance \cite{Pissondes}
with regard to real motion.
If one reverses the sign of time in the proper time differential element, the
velocity $\vec{v}_+$ becomes $\vec{v}_-$ and there is no reason for these
velocities to be equal, in contrast to what happens in the standard case.
The quadratic relativistic invariant of the special relativity is thus not
conserved. Let us consider violation of the reflection invariance caused
by the structural disparity characterized with the structural velocity $\vec{u}$.
We insist simultaneously on the requirement, that breaking of the reflection
invariance does not disturb spatial isotropy.
This can be achieved by the transition to a new structural reference system $S$ by the
relation
\begin{equation}
\vec{r}_2 = \vec{r}, \ \ \ \ \ \
t_2 = t - \vec{u}\!\cdot\!\vec{r}.
\label{eq:x51}
\end{equation}
The reference system $S$ is associated with the fractal structure of the
object $B$ which interacts with the fractal object $A$.
The interaction of both fractals makes substantial distinction between the systems
$S$ and $S_2$.
In the same way one can introduce the coordinate system connected with
the fractal structure of the object $A$ which interacts with the fractal $B$.
The only difference is the interchange $\vec{u}\leftrightarrow -\vec{u}$.
Let us examine the reference system $S$ in more detail.
Distortion of space-time in this system is given by the metrics
\begin{equation}
\eta(\vec{u}) =
\left(
\begin{array}{cc}
-\delta_{ij}+ u_iu_j & -u_i \\
-u_j & 1 \\
\end{array}
\right)
\label{eq:x52}
\end{equation}
which corresponds to the invariant
\begin{equation}
t^2-r^2-2t\vec{u}\!\cdot\!\vec{r} + (\vec{u}\!\cdot\!\vec{r})^2 =\tau^2.
\label{eq:x53}
\end{equation}
Simple metrics of this type has been used in the 3+1 formalism of general
relativity \cite{Misner}. In this formalism, space-time is described as a foliation
of space-like hyper-surfaces of constant time t. The quantity $\vec{u}$ has meaning
of a vector relating the spatial coordinate systems on different hyper-surfaces.
In difference from the structural velocity $\vec{u}$, we denote the motion velocity
as
\begin{equation}
\vec{v} = \frac{d\vec{r}}{dt}.
\label{eq:x54}
\end{equation}
Using the four-dimensional notation $r=(\vec{r},t)$,
the relativistic transformations with respect to motion
which preserve the invariant (\ref{eq:x53}) can be expressed as follows
\begin{eqnarray}
r'' = \Delta_D(\vec{v},\vec{u})r',
\label{eq:x55}
\end{eqnarray}
where
\begin{equation}
\Delta_D(\vec{v},\vec{u})=D^{-1}(\vec{u})\Lambda(\vec{v}_2)D(\vec{u}).
\label{eq:x56}
\end{equation}
The $\Lambda(\vec{v}_2)$ is Lorenz transformation matrix of the form (\ref{eq:x39})
which depends on the motion velocity vector
\begin{equation}
\vec{v}_2 = \frac{\vec{v}}{1-\vec{u}\!\cdot\!\vec{v}}
\label{eq:x57}
\end{equation}
and
\begin{equation}
D(\vec{u}) =
\left(
\begin{array}{cc}
\delta_{ij} & 0 \\
-u_j & 1 \\
\end{array}
\right) .
\label{eq:x58}
\end{equation}
The transformation matrix $\Delta_D$ can be rewritten into the compact form
\begin{equation}
\Delta_D(\vec{v},\vec{u}) =
\left(
\begin{array}{cc}
\delta_{ij}\!+\!G v_iv_j\!+\!\Gamma v_iu_j &
-\Gamma v_i \\
-G_{-}v_j\!-\!\Gamma_{-}u_j & 1\!+\!\Gamma_{-} \\
\end{array}
\right) .
\label{eq:x59}
\end{equation}
Here we have used the notations
\begin{equation}
\Gamma = \frac{1}
{\sqrt{(1-\vec{u}\!\cdot\!\vec{v})^2 - v^2}} ,
\label{eq:x60}
\end{equation}
\begin{equation}
G = \frac{(1-\vec{u}\!\cdot\!\vec{v})\Gamma-1}{v^2}
\label{eq:x61}
\end{equation}
and
\begin{equation}
\Gamma_{\pm} = Gv^2 \pm \Gamma\vec{u}\!\cdot\!\vec{v} ,
\ \ \ \ \ \ \ \
G_{\pm} =
\Gamma \pm G\vec{u}\!\cdot\!\vec{v} .
\label{eq:x62}
\end{equation}
The factor $\Gamma$ is analogue of the Lorenz factor for non-zero space-time
anisotropy $\vec{u}$ relating particle's
proper time $\tau$ with the time $t$ in the reference system $S$,
\begin{equation}
t=\tau\Gamma .
\label{eq:x63}
\end{equation}
The transformations inverse to (\ref{eq:x55}) are obtained by the interchange
$r'' \leftrightarrow r'$ and replacing $\vec{v}$ by $\vec{v}_{inv}$,
where
\begin{equation}
\vec{v}_{inv} = -\frac{\vec{v}}{1-2\vec{u}\!\cdot\!\vec{v}} .
\label{eq:x64}
\end{equation}
This formula connects the motion velocity $\vec{v}$ of a system $S'$ in the reference
system $S$ with the motion velocity $\vec{v}_{inv}$ of the system $S$ in the $S'$
reference frame. Because of space-time structural anisotropy $\vec{u}$,
the magnitudes of the two motion velocities are not equal.
Exploiting the symmetry properties
\begin{eqnarray}
\Gamma(\vec{v}_{inv}) =
(1-2\vec{u}\!\cdot\!\vec{v})\Gamma(\vec{v}) , \ \ \ \ \
\Gamma_{\pm}(\vec{v}_{inv}) = \Gamma_{\mp}(\vec{v}) ,
\nonumber \\
G(\vec{v}_{inv}) = (1-2\vec{u}\!\cdot\!\vec{v})^2G(\vec{v}) , \ \ \ \ \ \
G_{\pm}(\vec{v}_{inv}) =
(1-2\vec{u}\!\cdot\!\vec{v})G_{\mp}(\vec{v}) ,
\label{eq:x65}
\end{eqnarray}
the transformation matrix of the inverse transformations reads
\begin{equation}
\Delta_D^{-1}(\vec{v},\vec{u}) =
\left(
\begin{array}{cc}
\delta_{ij}\!+\!G v_iv_j\!-\!\Gamma v_iu_j &
+\Gamma v_i \\
+G_{+}v_j\!-\!\Gamma_{+}u_j & 1\!+\!\Gamma_{+} \\
\end{array}
\right) .
\label{eq:x66}
\end{equation}
As follows from the relation
\begin{equation}
\Delta_D^{\dag}(\vec{v},\vec{u}) \eta(\vec{u}) \Delta_D(\vec{v},\vec{u}) =
\eta(\vec{u}) ,
\label{eq:x67}
\end{equation}
the motion transformations (\ref{eq:x55}) preserve the invariant (\ref{eq:x53}).
The transformations comply the principle of relativity expressed by their group properties.
The composition of the  transformations has the form
\begin{equation}
\Omega_D(\vec{\phi},\vec{u}) \Delta_D(\vec{v}'',\vec{u})  =
\Delta_D(\vec{v}',\vec{u}) \Delta_D(\vec{v},\vec{u})
\label{eq:x68}
\end{equation}
with $\Omega_D(\vec{\phi},\vec{u})=D^{-1}(\vec{u}) R(\vec{\phi}) D(\vec{u})$,
provided
\begin{equation}
\vec{v}'' = \frac{\vec{v}' +
\vec{v}\left[\Gamma(1\!-\!\vec{u}\!\cdot\!\vec{v}')
+G\vec{v}\!\cdot\!\vec{v}'\right]  }
{1+\Gamma_{+}(1\!-\!\vec{u}\!\cdot\!\vec{v}')
+G_{+}\vec{v}\!\cdot\!\vec{v}'} .
\label{eq:x69}
\end{equation}
The inverse relation reads
\begin{equation}
\vec{v}' = \frac{\vec{v}'' -
\vec{v}\left[\Gamma(1\!-\!\vec{u}\!\cdot\!\vec{v}'')
-G\vec{v}\!\cdot\!\vec{v}''\right]  }
{1+\Gamma_{-}(1\!-\!\vec{u}\!\cdot\!\vec{v}'')
-G_{-}\vec{v}\!\cdot\!\vec{v}''} .
\label{eq:x70}
\end{equation}
One can obtain the above relations from the standard composition of the
$\Lambda$ matrices.
Range of the accessible values of the velocities $\vec{v}$ is given by the rotational
ellipsoid
\begin{equation}
(v_{\parallel}+e)^2 + \gamma^2v^2_{\bot} = \gamma^4.
\label{eq:x71}
\end{equation}
Here $v_{\parallel}$ and $v_{\bot}$ denote the velocity components
which are parallel and perpendicular to the space-time structural anisotropy
$\vec{u}$, respectively. The ellipsoid is given by the major semi-axis
$a=\gamma^2$ and by the minor semi-axis $b=\gamma$. Its eccentricity
is $e=\gamma\sqrt{\gamma^2\!-\!1}$.
One focus of the ellipsoid corresponds to the point $\vec{v}=0$.
The velocity ellipsoid is invariant with respect to the
relations (\ref{eq:x69}) and (\ref{eq:x70}).

In the case of $\vec{v}=(0,0,v)$ and $\vec{u}=(0,0,u)$, the
motion relativistic transformations have the simple form
\begin{eqnarray}
r''_z = \Gamma[(1-2uv)r'_z+vt'], \ \ \ \ \
t'' = \Gamma[t'+vr'_z\gamma^{-2}]
\label{eq:x72}
\end{eqnarray}
\begin{eqnarray}
r'_z = \Gamma[r''_z-vt''], \ \ \ \ \ \
t' = \Gamma[(1-2uv)t''-vr''_z\gamma^{-2}] .
\label{eq:x73}
\end{eqnarray}
The transverse components are conserved, $r''_i=r'_i, i = x,y$.
Composition of the corresponding velocities reads
\begin{equation}
v''_z =
\frac{v'_z+v-2u vv'_z}{1+vv'_z\gamma^{-2}}, \ \ \
v''_{\bot} = v'_{\bot}\frac{\Gamma^{-1}}{1+vv'_z\gamma^{-2}} ,
\label{eq:x74}
\end{equation}
\begin{equation}
v'_z =
\frac{v''_z-v}{1-2uv-vv''_z\gamma^{-2}}, \ \ \
v'_{\bot} = v''_{\bot}\frac{\Gamma^{-1}}{1-2uv-vv''_z\gamma^{-2}} .
\label{eq:x75}
\end{equation}
Detailed classification of the linear transformations of the type
(\ref{eq:x72}) and (\ref{eq:x73}) was performed in 1+1 dimensions in Ref.
\cite{Lalan}.

\vskip 0.5cm
{\section {Relativistic mechanics in anisotropic space-time}}

In standard relativistic mechanics, the position and momentum of an elementary
particle is given by the four-vectors
$r^{\mu}=\{\vec{r},t\}$ and $p^{\mu}=\{\vec{p},E\}$,
respectively.
We comprehend the notion of elementarity as a relative
concept which relies on the scales and structures we are dealing with.
For the infinite resolution the elementary particle should be a perfect
point without any internal structure.
For an arbitrary small but still finite resolution,
the perfect point is approximated by a particle which we
call "elementary" with respect to this resolution.
It is therefore natural to assume that
the concepts of the momentum, energy,  mass and the velocity of the
"elementary" particle have good physical meaning also
in the region where the space-time isotropy is violated.
Let us consider space-time structural anisotropy
induced by collisions of two interacting fractals.
Suppose the anisotropy is characterized by the structural velocity $\vec{u}$.
In such case we have to impose general requirements
on mechanical variables, which remain still valid.
Based on our physical intuition, we formulate these requirements as follows:

1. Energy of a free particle cannot be pumped from the
structure of space-time. This condition reads
\begin{equation}
E = E_{min} = m_0 \ \ \ \ \ \ for \ \ \ \ \ \vec{v}=0
\label{eq:x76}
\end{equation}
where $m_0$ is the rest mass and $\vec{v}$ is the velocity of the free particle.

2. Rate of clocks is slowest in the centre of gravity system.
The only source of gravity is free particle itself and
not the structure of space-time. This condition reads
\begin{equation}
dt = dt_{min} \ \ \ \ \ \ for \ \ \ \ \ \vec{P}=0
\label{eq:x77}
\end{equation}
where the vector $\vec{P}$ defines effect of force on the particle.
The physical requirements are obvious in the Minkovski space-time. In the
case of the space-time anisotropy, they lead to specific constraints on
the mechanical variables. The construction depends on the way the anisotropy is
induced by the interaction.
It will be instructive to discuss the situation corresponding to the
transformations (\ref{eq:x35}) and (\ref{eq:x43}) separately.

\vskip 0.5cm
{\subsection {Fractal interaction with torsion}}

The transformations with structural torsion model situation when the interacting
fractal structures are in mutual spinning position.
Amount of the mutual torsion depends on the values of anomalous
fractal dimensions of the colliding fractal objects which can be functions of their
spin states. Explicit spin dependencies of these parameters have to be determined
from experiment and require further independent study.
Without going to details, we show here that the phenomenological aspects of the
requirements (\ref{eq:x76}) and (\ref{eq:x77}) can be fulfilled
in the following way.
Relying on the transformations (\ref{eq:x43}), we link
the mechanical variables $\vec{p}_2$ and $E_2$ defined in the isolated fractal
reference frame $S_2$ with their counterparts $\vec{P}$ and $E$ in the
reference system $S$ by the expressions
\begin{equation}
\vec{p}_2 = \gamma^{-1}\vec{P}-\sigma\vec{u}\!\times\!\vec{P}-\vec{u}E , \ \ \ \ \
E_2 = E .
\label{eq:x78}
\end{equation}
The $\vec{u}$ is the structural velocity considered in the previous sections.
Last equation suggests that energy of a free particle is the same in the isolated fractal
system $S_2$ and the reference frame $S$ which is essentially the same fractal system
disturbed by the interaction with the fractal object $A$.
This is not longer true for the quantities $\vec{p}_2$ and $\vec{P}$.
The relation between them which is inverse to Eq. (\ref{eq:x78}) reads
\begin{equation}
\vec{P} =
\gamma^{-1}\vec{p}_2+\sigma\vec{u}\!\times\!\vec{p}_2+
\gamma\vec{u}(E_2+\vec{u}\!\cdot\!\vec{p}_2) .
\label{eq:x79}
\end{equation}
The standard relativistic invariant in the system $S_2$
is replaced by the invariant expression
\begin{equation}
\gamma^{-2}[E^2-P^2+2E\vec{U}\!\cdot\!\vec{P} -(\vec{U}\!\times\!\vec{P})^2] = m_0^2 .
\label{eq:x80}
\end{equation}
This corresponds to the the metrics
\begin{equation}
\bar{\eta}(\vec{u}) =
\left(
\begin{array}{cc}
-\delta_{ij}+ u_iu_j & u_i\gamma^{-1} \\
u_j\gamma^{-1} & \gamma^{-2} \\
\end{array}
\right)
\label{eq:x81}
\end{equation}
in the $(\vec{P},E)$ space.
The relativistic transformations of the mechanical variables with respect to motion,
which preserve the invariant (\ref{eq:x80}),
can be written as follows
\begin{equation}
P'' = \Delta_H(\vec{W},\vec{U})P'
\label{eq:x82}
\end{equation}
where
\begin{equation}
\Delta_H(\vec{W},\vec{U}) = H^{-1}(\vec{u})\Lambda(\vec{v}_2)H(\vec{u}).
\label{eq:x83}
\end{equation}
Here $\Lambda(\vec{v}_2)$ has the same form as in Eq. (\ref{eq:x39}) and
\begin{equation}
H(\vec{u}) =
\left(
\begin{array}{cc}
\gamma^{-1}\delta_{ij}+\sigma\epsilon_{ijk}u_k & -u_i \\
0 & 1 \\
\end{array}
\right) .
\label{eq:x84}
\end{equation}
When evaluating the right-hand side of the expression (\ref{eq:x83}) it can be shown that
the transformation matrix $\Delta_H$ is explicit function of the anisotropy
$\vec{U}$ and the vector
\begin{equation}
\vec{W} = \gamma^{-1}\vec{v}+\sigma\vec{u}\!\times\!\vec{v} .
\label{eq:x85}
\end{equation}
It takes the form
\begin{equation}
\Delta_H(\vec{W},\vec{U}) =
\left(
\begin{array}{cc}
\delta_{ij}\!+\!G W_iW_j\!-\!G_{-} U_iW_j &
-G_{+}W_i + GW^2U_i\\
-\Gamma W_j & 1\!+\!\Gamma_{+} \\
\end{array}
\right)
\label{eq:x86}
\end{equation}
where the symbols $\Gamma$, $G$, $\Gamma_{\pm}$, and $G_{\pm}$  are given by
Eqs. (\ref{eq:x60}), (\ref{eq:x61}), and (\ref{eq:x62}), respectively.
They can be expressed in terms of $\vec{U}$ and $\vec{W}$ in the following way
\begin{equation}
\Gamma = \frac{1}
{\sqrt{1-2\vec{U}\!\cdot\!\vec{W} - W^2}} , \ \ \ \ \ \ \
G = \frac{(1-\vec{U}\!\cdot\!\vec{W})\Gamma-1}{W^2+(\vec{U}\!\cdot\!\vec{W})^2},
\label{eq:x87}
\end{equation}
\begin{equation}
\Gamma_{\pm} = GW^2+G(\vec{U}\!\cdot\!\vec{W})^2 \pm \Gamma\vec{U}\!\cdot\!\vec{W} ,
\ \ \ \ \ \ \ \
G_{\pm} =
\Gamma \pm G\vec{U}\!\cdot\!\vec{W} .
\label{eq:x88}
\end{equation}
The transformations inverse to Eq. (\ref{eq:x82}) are obtained by the interchange
$P'' \leftrightarrow P'$ and replacing  $\vec{W}$ by $\vec{W}_{inv}$,
where
\begin{equation}
\vec{W}_{inv} = -\frac{\vec{W}}{1-2\vec{U}\!\cdot\!\vec{W}} .
\label{eq:x89}
\end{equation}
Exploiting the symmetry properties (\ref{eq:x65}),
the inverse transformation matrix reads
\begin{equation}
\Delta_H^{-1}(\vec{W},\vec{U}) =
\left(
\begin{array}{cc}
\delta_{ij}\!+\!G W_iW_j\!+\!G_{+} U_iW_j &
G_{-}W_i + GW^2U_i\\
\Gamma W_j & 1\!+\!\Gamma_{-} \\
\end{array}
\right) .
\label{eq:x90}
\end{equation}
As follows from the relation
\begin{equation}
\Delta_H^{\dag}(\vec{W},\vec{U}) \bar{\eta}(\vec{u}) \Delta_H(\vec{W},\vec{U}) =
\bar{\eta}(\vec{u}) ,
\label{eq:x91}
\end{equation}
the motion transformations of mechanical variables preserve the invariant (\ref{eq:x80}).
The composition of the  transformations has the form
\begin{equation}
\Omega_H(\vec{\phi},\vec{U}) \Delta_H(\vec{W}'',\vec{U})  =
\Delta_H(\vec{W}',\vec{U}) \Delta_H(\vec{W},\vec{U})
\label{eq:x92}
\end{equation}
with $\Omega_H(\vec{\phi},\vec{U})=H^{-1}(\vec{u}) R(\vec{\phi}) H(\vec{u})$,
provided
\begin{equation}
\vec{W}'' = \frac{\vec{W}' +
\vec{W}\left[\Gamma\!-\!G_{-}\vec{U}\!\cdot\!\vec{W}'+G\vec{W}\!\cdot\!\vec{W}'\right]  }
{1+\Gamma_{+}\!-\!GW^2\vec{U}\!\cdot\!\vec{W}'+G_{+}\vec{W}\!\cdot\!\vec{W}'} .
\label{eq:x93}
\end{equation}
The inverse formula reads
\begin{equation}
\vec{W}' = \frac{\vec{W}'' -
\vec{W}\left[\Gamma\!-\!G_{+}\vec{U}\!\cdot\!\vec{W}''-G\vec{W}\!\cdot\!\vec{W}''\right]  }
{1+\Gamma_{-}\!-\!GW^2\vec{U}\!\cdot\!\vec{W}''-G_{-}\vec{W}\!\cdot\!\vec{W}''} .
\label{eq:x94}
\end{equation}
The above relations can be obtained from the standard composition of the $\Lambda$
matrices.
We end up this section by the following observation.
White the transformations of the kinematical variables depend on the quantities
$\vec{u}$ and $\vec{v}$, the transformations of the mechanical variables are explicit
functions of $\vec{U}$ and $\vec{W}$.
The variables $r^{\mu}$ and $P^{\mu}$ posses different
transformation properties with respect to motion.
The first obey the transformation formula (\ref{eq:x55}),
the later are transformed according to Eq. (\ref{eq:x82}).
This separation of the kinematical and mechanical
sector is characteristic property for space-time  with non-zero anisotropy.

\vskip 0.5cm
{\subsection {Fractal interaction without torsion}}

The transformations without structural torsion model situation when the interacting fractal
structures do not spin mutually around the collision axis.
In such case, the requirements (\ref{eq:x76}) and (\ref{eq:x77})
have to be fulfilled in consistence with the structural transformations (\ref{eq:x35}).
We connect therefore the mechanical variables $\vec{p}_2$ and $E_2$ defined in the
isolated fractal reference frame $S_2$ with their counterparts $\vec{P}$ and $E$
in the reference system $S$ by the relations
\begin{equation}
\vec{p}_2 =
\vec{P}-\vec{u}\frac{\vec{u}\!\cdot\!\vec{P}}{u^2}
\left(\gamma^{-1}-1\right)
-\vec{u}E , \ \ \ \ \
E_2 = E .
\label{eq:x95}
\end{equation}
Also here we require equality of the energy of a free particle in the
system $S_2$ with its value in the reference frame $S$
being the same fractal system disturbed by the interacting fractal object $A$.
The relation between $\vec{p}_2$ and $\vec{P}$ differs however from Eq. (\ref{eq:x78}).
The inverse relation reads
\begin{equation}
\vec{P} =
\vec{p}_2+\vec{U}\frac{\vec{U}\!\cdot\!\vec{p}_2}{U^2}
\left(\gamma-1\right)
+\vec{U}E_2 .
\label{eq:x96}
\end{equation}
The standard relativistic invariant in the system $S_2$ is replaced
by the invariant expression
\begin{equation}
\gamma^{-2}(E+\vec{U}\!\cdot\!\vec{P})^2 -P^2 = m_0^2
\label{eq:x97}
\end{equation}
corresponding to the metrics
\begin{equation}
\bar{\eta}(\vec{u}) =
\left(
\begin{array}{cc}
-\delta_{ij}+ u_iu_j & u_i\gamma^{-1} \\
u_j\gamma^{-1} & \gamma^{-2} \\
\end{array}
\right) .
\label{eq:x98}
\end{equation}
The relativistic invariant and the metrics in $(\vec{P},E)$ space are
thus identical for both cases, the interaction with and without torsion.
The same is true for the corresponding relativistic transformations with respect to motion.
Really, from the relations (\ref{eq:x95}) we see that the transformations
can be written as follows
\begin{equation}
P'' = \Delta_{\tilde{H}}(\vec{W},\vec{U})P'
\label{eq:x99}
\end{equation}
where
\begin{equation}
\Delta_{\tilde{H}}(\vec{W},\vec{U}) =
\tilde{H}^{-1}(\vec{u})\Lambda(\vec{v}_2)\tilde{H}(\vec{u}) .
\label{eq:x100}
\end{equation}
Here $\Lambda(\vec{v}_2)$ is the same as in Eq. (\ref{eq:x83}) while
\begin{equation}
\tilde{H}(\vec{u}) =
\left(
\begin{array}{cc}
\delta_{ij}+(\gamma^{-1}\!-\!1)u_iu_j/u^2      & -u_i \\
0 & 1 \\
\end{array}
\right).
\label{eq:x101}
\end{equation}
When evaluating the right-hand side of the expression (\ref{eq:x100}), it can be shown that
the transformation matrix $\Delta_{\tilde{H}}$ is explicit function of the anisotropy
$\vec{U}$ and the vector
\begin{equation}
\vec{W} = \vec{v} + \vec{u}\frac{\vec{u}\!\cdot\!\vec{v}}{u^2}
\left(\gamma^{-1}-1\right) .
\label{eq:x102}
\end{equation}
It can be rewritten to the form
\begin{equation}
\Delta_{\tilde{H}}(\vec{W},\vec{U}) =
\left(
\begin{array}{cc}
\delta_{ij}\!+\!G W_iW_j\!-\!G_{-} U_iW_j &
-G_{+}W_i + GW^2U_i\\
-\Gamma W_j & 1\!+\!\Gamma_{+} \\
\end{array}
\right)
\label{eq:x103}
\end{equation}
which is identical with the matrix (\ref{eq:x86}), $\Delta_{\tilde{H}}\equiv\Delta_H$.
The $\Gamma$, $G$, $\Gamma_{\pm}$, and  $G_{\pm}$ are the same functions of
the quantity $\vec{W}$ (given here by Eq. (\ref{eq:x102})) as the expressions
(\ref{eq:x60}), (\ref{eq:x61}), and (\ref{eq:x62}), respectively. The identical
relations as in the previous subsection
are valid for the inverse transformations and for
the composition rules of $\vec{W}$.

We again observe, that the transformations of the mechanical
variables with respect to motion are different from the transformations
of the kinematical variables.
The mechanical sector characterized by the quantities $\vec{U}$ and $\vec{W}$
is identical with the mechanical sector for the interactions with torsion.
The only difference is way the quantity $\vec{W}$ is linked to the
kinematical velocity $\vec{v}$.
While for interactions with structural torsion this relation is given
by Eq. (\ref{eq:x85}), for the interactions without torsion it follows from
Eq. (\ref{eq:x102}).

\vskip 0.5cm
{\section {Relations of the kinematical and mechanical
variables}}

Fundamental concepts of the special theory of relativity lead us
to the relation between the energy/momentum of a
particle and its velocity. The velocity is limited within the sphere
of the radius $c=1$ in every inertial system of reference and is oriented in
the direction of the particle momentum. Coefficient of the proportionality
between the momentum and the velocity is relativistic mass of the particle.
The relativistic mass is equivalent to the particle's energy.
Change of the particle's momentum per unit time defines force the
particle is acted upon. This concerns the homogeneous and isotropic space-time.
We show how the relations modify provided structural violation of
the space-time isotropy characterized by the structural velocity $\vec{u}$.

We start with the relations (\ref{eq:x45}) valid in an isolated fractal
reference frame $S_2$. The corresponding relations in the reference system $S$
are different. Exploiting Eqs. (\ref{eq:x51}) and (\ref{eq:x78}), or
(\ref{eq:x51}) and (\ref{eq:x95}) we get
\begin{equation}
\vec{P} = M(\vec{W}+\vec{U}), \ \ \ \ \ \ \
E = M(1-\vec{U}\!\cdot\!\vec{W})
\label{eq:x104}
\end{equation}
where
\begin{equation}
M = \Gamma m_0, \ \ \ \ \ \ \
\Gamma = \frac{1}{\sqrt{1-2\vec{U}\!\cdot\!\vec{W}-W^2}}.
\label{eq:x105}
\end{equation}
Form of these expressions is the same for fractal interaction
with or without torsion, respectively. The $\vec{W}$ depends, however,
in both cases on the motion velocity $\vec{v}$ in a different way.
The expressions (\ref{eq:x104}) are form invariant with respect to motion.
This allows to consider the quantity $M$ as meaningful generalization
of the relativistic mass of a free particle in space-time with the
structural anisotropy $\vec{u}$.
We see from the above relations that the energy $E$ and the mass $M$ become
independent. The energy is always larger than the particle's rest mass $m_0$, $E\ge m_0$.
Minimum of the energy is acquired for $\vec{W}=0$  and thus for the non-zero value
of $\vec{P}=\vec{P}_0$,
\begin{equation}
E_{min} = E(\vec{P}_0) = m_0  , \ \ \ \ \ \ \ \
\vec{P}_0 = m_0\vec{U} = \frac{m_0\vec{u}}{\sqrt{1\!-\!u^2}} .
\label{eq:x106}
\end{equation}
This is in consistence with the requirement (\ref{eq:x76}), because
for both cases, the fractal interaction with and/or without torsion,
Eqs. (\ref{eq:x85}) and (\ref{eq:x102}) give the zero motion velocity $\vec{v}=0$
when $\vec{W}=0$.
Contrary to the energy $E$, the mass $M$ of a particle can be even smaller
than its rest mass $m_0$.
The minimal value of particle's mass is acquired for
$\vec{P}=0$  and thus for the non-zero value of $\vec{W}=\vec{W}_0$,
\begin{equation}
M_{min} = M(\vec{W}_0) \le m_0 , \ \ \ \ \ \ \ \ \ \
\vec{W}_0 = -\vec{U}.
\label{eq:x107}
\end{equation}
This corresponds to the minimum of the factor $\Gamma$ and to the minimal
value of t (\ref{eq:x63}). Consequently, the requirement (\ref{eq:x77}) is
thus fulfilled as well.
The minimal mass depends on the value of the space-time anisotropy. Formal relation between
the rest mass $m_0$ and the minimal mass $M_{min}$ can be written
in the form
\begin{equation}
m_0 = \gamma M_{min}.
\label{eq:x108}
\end{equation}
From the above relations we conclude that, in space-time with structural
anisotropy,  the energy and mass of a free particle become independent quantities.

We shall demonstrate this in a more formal way.
The energy $E$ of a particle with the rest mass $m_0$
is function of two independent quantities, $\vec{U}$ and $\vec{P}$.
This can be obtained when solving the invariant (\ref{eq:x80}) or (\ref{eq:x97})
with respect to the $E$. One gets
\begin{equation}
E(\vec{U},\vec{P}) = \sqrt{1\!+\!U^2}\sqrt{P^2+m^2_0} - \vec{U}\!\cdot\!\vec{P} .
\label{eq:x109}
\end{equation}
Note that this relation, when expressed in terms of $u$,
\begin{equation}
E = \gamma\left(\sqrt{P^2+m^2_0} - \vec{u}\!\cdot\!\vec{P}\right),
\label{eq:x110}
\end{equation}
represents the relativistic transformation of energy between
the isolated fractal reference frames $S_1$ and $S_2$
(see Eqs. (\ref{eq:x35}) and (\ref{eq:x43})).
When calculating partial derivatives of the energy $E(\vec{U},\vec{P})$
with respect to $\vec{P}$ and $\vec{U}$, one obtains
\begin{equation}
\frac{\partial E}{\partial P_i} = W_i , \ \ \ \ \ \ \ \
\frac{\partial E}{\partial U_i} = -MW_i .
\label{eq:x111}
\end{equation}
Because $U_i$ are space components of the space-time structural four-velocity,
we can consider the above two partial differential equations as an analogue
of the Hamilton equations.
First of them serves as definition of $\vec{W}$. In this way the vector $\vec{W}$ is
defined as partial derivative which involves the mechanical variables $E$ and
$\vec{P}$. Therefore, we will refer to it as "mechanical velocity". By this
name we want to distinguish the mechanical velocity $\vec{W}$ from the
"kinematical velocity" $\vec{v}$ (\ref{eq:x54}) which is defined by means of
pure kinematical variables $\vec{r}$ and $t$.
Unlike the structural velocity $\vec{u}$, both kinematical and mechanical
velocity are quantities reflecting amount of motion.
The explicit relations between them
(Eqs. (\ref{eq:x85}) and (\ref{eq:x102})) depend on the
way the space-time anisotropy is induced by the interaction.

The second equation in (\ref{eq:x111}) can be exploited by the independent definition
of the particle mass. We can define namely
\begin{equation}
M = - \frac{\partial E}{\partial U_i}
\left(\frac{\partial E}{\partial P_i}\right)^{-1}.
\label{eq:x112}
\end{equation}
Using this definition and exploiting Eq. (\ref{eq:x109}), we obtain the formula
\begin{equation}
M(U,P) = \sqrt{\frac{P^2+m^2_0}{1+U^2}}
\label{eq:x113}
\end{equation}
which depends on the magnitudes of both $U$ and $P$.
Inserting this expression into the first equation (\ref{eq:x111}), we get
\begin{equation}
\frac{\vec{P}}{M}-\vec{U} = \vec{W} .
\label{eq:x114}
\end{equation}
From here the relations (\ref{eq:x104}) and (\ref{eq:x105}) follow immediately.
When considering the vector $\vec{U}$ as a scale dependent fluctuating
anisotropy parameter of space-time, the mass of a particle can be treated as
a quantity proportional to the change of particle's energy with these
space-time fluctuations.
At small scales the characteristic size of fluctuations increases and the mass decreases.
On the contrary when the resolution decreases, the fluctuations become negligible.
For small $\vec{U}\rightarrow 0$ the change of energy with fluctuations remains
finite
\begin{equation}
\frac{\partial E}{\partial U_i} \rightarrow -p_i,
\label{eq:x115}
\end{equation}
allowing for smooth limit
\begin{equation}
M \rightarrow  \frac{m_0}{\sqrt{1\!-\!v^2}}.
\label{eq:x116}
\end{equation}
The independence of the energy $E$ and the mass $M$ of a particle is only one of the
consequences in the anisotropic space-time.
For the same reasons we have to distinguish the particle's momentum
\begin{equation}
\vec{p} \equiv  M\vec{v}
\label{eq:x117}
\end{equation}
from the "impulse" $\vec{P}$ of the particle standing in the text in upper case
notation. The momentum $\vec{p}$ (or more precisely the kinematical momentum)
is product of the particle's mass $M$ and its kinematical velocity $\vec{v}$.
It satisfies the standard dispersion relation  between the
energy $E$ and the rest mass $m_0$,
\begin{equation}
E^2 = (Mv)^2 + m^2_0 .
\label{eq:x118}
\end{equation}

In classical mechanics, when the velocity of a particle
and therefore its momentum are constant in time, this indicates that the
particle is free. If, however, the momentum of the particle changes with time,
the particle is said to be acted upon by a force.
In anisotropic space-time force, work and the kinetic energy is directly connected
with the "impulse" $\vec{P}$ of the particle.
The force acting on the particle is equal to change of the particle's "impulse"
$\vec{P}$ per unit time,
\begin{equation}
\vec{F} =  \frac{d\vec{P}}{dt} .
\label{eq:x119}
\end{equation}
The connection between $\vec{F}$ defined in
space-time with structural anisotropy $\vec{u}$ and the force $\vec{F}_2$ expressed in
the isolated fractal reference system $S_2$ reads
\begin{equation}
\vec{F}_2 \equiv  \frac{d\vec{p}_2}{dt_2} =
\frac{1}{(1\!-\!\vec{u}\!\cdot\!\vec{v})}\left(
\vec{F}\gamma^{-1}-\sigma\vec{u}\!\times\!\vec{F}-\vec{u}\frac{dE}{dt}\right)
\label{eq:x120}
\end{equation}
or
\begin{equation}
\vec{F}_2 \equiv  \frac{d\vec{p}_2}{dt_2} =
\frac{1}{(1\!-\!\vec{u}\!\cdot\!\vec{v})}\left(
\vec{F}+\vec{u}\frac{\vec{u}\!\cdot\!\vec{F}}{u^2}(\gamma^{-1}-1)
-\vec{u}\frac{dE}{dt}\right) ,
\label{eq:x121}
\end{equation}
in dependence the fractal interaction is with or without torsion, respectively.
In standard mechanics, the work $A$ done by a force per unit time is defined
as scalar product of the force and velocity.
This implies the same definition in any isolated fractal reference frame, in
particular in $S_2$,
\begin{equation}
A_2 = \vec{v}_2\!\cdot\!\vec{F}_2 .
\label{eq:x122}
\end{equation}
On the other side, the work $A_2$ equals to change of the
kinetic energy $T_2=E_2-m_0$ per unit time,
\begin{equation}
A_2 = \frac{dT_2}{dt_2}=\frac{dE_2}{dt_2} .
\label{eq:x123}
\end{equation}
Realizing that
\begin{equation}
\frac{dE_2}{dt_2} = \frac{1}{1\!-\!\vec{u}\!\cdot\!\vec{v}}\frac{dE}{dt}  ,
\label{eq:x124}
\end{equation}
and exploiting  Eqs. (\ref{eq:x120}) or (\ref{eq:x121}), we get
\begin{equation}
\frac{dE}{dt} = \vec{W}\!\cdot\!\vec{F} ,
\label{eq:x125}
\end{equation}
where $\vec{W}$ is given by Eqs. (\ref{eq:x85}) or (\ref{eq:x102}), respectively.
This relation states that, in space-time with the structural anisotropy $\vec{u}$,
the change of energy per unit time is equal to the scalar product of the
acting force $\vec{F}$ and the mechanical velocity $\vec{W}$.
Inserting here the expression (\ref{eq:x119}) for $\vec{F}$, we arrive again
at the first equation (\ref{eq:x111}).

One can proceed in the reverse order and define the kinetic energy $T$ by the
equation
\begin{equation}
\frac{dT}{dt} = A = \vec{W}\!\cdot\!\vec{F}
\label{eq:x126}
\end{equation}
which means that change of the kinetic energy per unit time is equal to the
work $A$. Using (\ref{eq:x119}) and (\ref{eq:x104}) the right-hand side of
Eq. (\ref{eq:x126}) may be rewritten to the form
\begin{equation}
A = \left[
\vec{W}\!\cdot\!\frac{d}{dt}
M(\vec{W}+\vec{U})
\right] =
M\left(\vec{W}\!\cdot\!\frac{d\vec{W}}{dt}\right)
+M\left(\vec{U}\!\cdot\!\frac{d\vec{W}}{dt}+W\frac{dW}{dt}\right)
\frac{(W^2+\vec{U}\!\cdot\!\vec{W})}
{(1\!-\!2\vec{U}\!\cdot\!\vec{W}\!-\!W^2)}
= \frac{d}{dt}\left[M(1-\vec{U}\!\cdot\!\vec{W})\right].
\label{eq:x127}
\end{equation}
Here we have used the identity
\begin{equation}
\vec{W}\!\cdot\!\frac{d\vec{W}}{dt} = W\frac{dW}{dt} .
\label{eq:x128}
\end{equation}
Inserting expression (\ref{eq:x127}) to the right-hand side of Eq. (\ref{eq:x126}) and integrating
over $t$, we obtain for the kinetic energy of a particle
\begin{equation}
T = \frac{m_0(1\!-\!\vec{U}\!\cdot\!\vec{W})}{\sqrt{1\!-\!2\vec{U}\!\cdot\!\vec{W}\!-\!W^2}}
-m_0 .
\label{eq:x129}
\end{equation}
The integration constant is chosen so that $T(\vec{W}=0)=0$.
For small values of $\vec{W}$ compared with unity,
we can make an expansion in terms of $W_i$. We get to a first
approximation
\begin{equation}
T = \frac{1}{2}m_0\left(W^2+(\vec{U}\!\cdot\!\vec{W})^2\right)
=\frac{1}{2}m_0v^2.
\label{eq:x130}
\end{equation}
Last equality between $W^2$ and $v^2$ follows from both Eqs. (\ref{eq:x85}) and (\ref{eq:x102})
simultaneously.
For small velocities, we have obtained standard expression for
the kinetic energy in terms of the kinematical velocity $v$.

\vskip 0.5cm
{\section {Space-time anisotropy and Klein-Gordon equation }}

The relations (\ref{eq:x104}) and (\ref{eq:x118}) between the kinematical  and
mechanical variables are form invariant with respect to the motion transformations
(\ref{eq:x55}) and (\ref{eq:x82}), or (\ref{eq:x55}) and (\ref{eq:x99}), respectively.
The transformations preserve the quadratic forms (\ref{eq:x53}) and (\ref{eq:x80}).
Another relation which is invariant under these transformations is action of
a free particle. The action in space-time with anisotropy characterized by the structural velocity $\vec{u}$
can be written in the form
\begin{equation}
S_{u} = -Et +\vec{r}\!\cdot\!\vec{P}\gamma^{-1}
+\sigma\vec{u}\!\cdot\!(\vec{r}\!\times\!\vec{P}) ,
\label{eq:x131}
\end{equation}
or
\begin{equation}
S_{u} = -Et +\vec{r}\!\cdot\!\vec{P}
+\frac{(\vec{r}\!\cdot\!\vec{u})(\vec{u}\!\cdot\!\vec{P})}{u^2}\left(\gamma^{-1}-1 \right)
\label{eq:x132}
\end{equation}
in dependence on type of the fractal interaction.
Using the mechanical velocities (\ref{eq:x85}) and (\ref{eq:x102}), we can
define the vectors $\vec{X}(\vec{r},\vec{u}) = \vec{W}t$.
In terms of $\vec{X}$, both expressions (\ref{eq:x131}) and (\ref{eq:x132})
read
\begin{equation}
S_{u} = -\tau m_0 = -Et + \vec{P}\!\cdot\!\vec{X}(\vec{r},\vec{u}) .
\label{eq:x133}
\end{equation}
Having determined the action, we re-examine the Klein-Gordon
equation for a free particle.
The corresponding d'Alambertian operator is modified in the metrics (\ref{eq:x52}) as  follows
\begin{equation}
\Box_{u} = \partial^{\dag}[\eta(\vec{u})]^{-1}\partial
\label{eq:x134}
\end{equation}
where
\begin{equation}
[\eta(\vec{u})]^{-1} =
\left(
\begin{array}{cc}
-\delta_{ij}    & -u_i \\
-u_j & 1\!-\!u^2 \\
\end{array}
\right) .
\label{eq:x135}
\end{equation}
Here  $\partial = (\vec{\partial},\partial_0) $ is four-derivative with respect to the
four-coordinates $r=(\vec{r},t)$.
If we introduce the covariant derivatives
\begin{equation}
D_0 = \partial_0, \ \ \ \ \ \ \ \
\vec{D} = \vec{\partial}+\vec{u}\partial_0 ,
\label{eq:x136}
\end{equation}
the explicit form of the operator (\ref{eq:x134}) can be written in the way
\begin{equation}
\Box_{u} = \partial_0^{2}-
\left(\vec{\partial}+\vec{u}\partial_0\right)^{2} \equiv D_0^2- D^2 .
\label{eq:x137}
\end{equation}
One can check validity of the canonical operator equations
\begin{eqnarray}
iD_0 \psi_{u} &=& E \psi_{u} ,
\nonumber \\
-i\vec{D} \psi_{u} &=& M\vec{v}\psi_{u}
\label{eq:x138}
\end{eqnarray}
for  the solution $\psi_{u} = \exp(iS_{u})$, where $S_{u}$ is given by Eq. (\ref{eq:x133}).
The corresponding modified Klein-Gordon equation
\begin{equation}
-\Box_{u} \psi_{u} = m_0^2\psi_{u}
\label{eq:x139}
\end{equation}
leads to the invariant relation (\ref{eq:x118})
between the energy $E$ and the (kinematical)
momentum $\vec{p}=M\vec{v}$ of a free particle with the rest mass $m_0$.
This relation remains valid for arbitrary non-zero space-time structural
velocity $\vec{u}$.
Invariance of the d'Alambertian operator and thus form-invariance
of the Klein-Gordon equation
with respect to the motion relativistic transformations (\ref{eq:x55})
is seen from the expression
\begin{equation}
\Box_{u} = \partial^{\dag}[\eta(\vec{u})]^{-1}\partial
= \partial^{'\dag}[\eta(\vec{u})]^{-1}\partial' = \Box'_{u} .
\label{eq:x140}
\end{equation}
Here we have exploited the decomposition
\begin{equation}
[\eta(\vec{u})]^{-1}= \Delta_D [\eta(\vec{u})]^{-1}\Delta_D^{\dag}
\label{eq:x141}
\end{equation}
and the corresponding transformation property
\begin{equation}
\partial' = (\Delta_D^{\dag})^{-1}\partial .
\label{eq:x142}
\end{equation}
The above relations reflect motion invariance of the Klein-Gordon equation expressed
in terms of the covariant derivatives. Form of the covariant derivatives
suggests to consider the space-time anisotropy $\vec{u}$ as a kind of calibration
symmetries inherent to the very structure of space-time. This should
include elementary quantum fields as possible source of the space-time fluctuations and
requires further study which lies beyond the scope of this work.

\vskip 0.5cm
{\section {Conclusions}}

The questions addressed in the paper were stimulated by fractal properties of the
$z$-scaling observed in the inclusive reactions at high energies.
In the present status of our investigations the fractality was considered with
respect to the constituent (parton) content of the colliding hadrons and nuclei.
We suppose that constituents of these extended objects are composed of smaller
constituents which in turn are built of even smaller constituents forming thus a structure
typical for fractals.
The scaling variable $z$ was constructed as a fractal measure
connecting kinematics of the constituent interactions with the anomalous fractal dimensions
$\delta_1$ and $\delta_2$ of the objects colliding at high energies.
Value of $z$ depends on resolution at which the underlying
interaction of constituents can be singled out of the reaction.
Insisting on the minimal possible resolution,
we have obtained relativistic transformations as functions of the resolution dependent
structural velocity $u$.
The generalization of these transformations to 3+1 dimensions includes two separate
situations; the interaction of fractals with and without mutual torsion.
The obtained transformations were interpreted as special
realization of structural relativity. Considered realization
of the relativistic principle was formulated with respect to the isolated structural
reference frames associated with the isolated fractal objects of various
anomalous dimensions.
More generally, in view of intrinsic relation between
the fractality of the interacting objects and the fractal structure
of space-time,  the isolated reference systems of structural relativity
have been considered as attributed to the very structure of the (QCD) vacuum as well.

Motivated by many investigations concerning fractal properties of space-time,
we have proceeded beyond the isolated fractal systems. This concerns our working hypothesis
that interaction of fractal objects with different anomalous fractal dimensions can induce
structural anisotropy of space-time.
We have demonstrated that this hypothesis leads to asymmetry between the relativistic
kinematics and relativistic mechanics.
The kinematical sector was parametrized by the four-coordinates
$r=(\vec{r}$, $t)$, the kinematical velocities $\vec{v}=d\vec{r}/dt$,
the structural velocities $\vec{u}$ and by their composition laws.
In the mechanical sector enter the energy $E$ of a particle, its impulse
$\vec{P}$, mechanical velocity $\vec{W}$ and their transformation properties which are
explicit functions of the structural anisotropy $\vec{U}$.
Due to such splitting, the space-time anisotropy would lead to separation of the particle's
mass $M$ from its energy $E$ and also to separation of the particle's mechanical
momentum $M\vec{W}$ from its impulse $\vec{P}$.
Both independent quantities, $M$ and $M\vec{W}$, where shown to enter into two equations
which are analogue of Hamiltonian equations in Newtonian mechanics.
Connection between the kinematical and mechanical sector is characterized by
the quantities such as kinematical momentum $\vec{p}=M\vec{v}$, force $\vec{F}=d\vec{P}/dt$,
work $A=\vec{W}\vec{F}$ done by the force by unit time, and the relation between
the kinematical and mechanical velocities $\vec{v}$ and $\vec{W}$.

In the considered special realization of space-time structural relativity,
the space-time structural anisotropy $\vec{U}$ (or the structural velocity $\vec{u}$)
was treated as a relative quantity governed by relativistic principles.
The formulation suggests  that space-time anisotropy can be induced in the
ultra-relativistic collisions of structural objects such as hadrons and nuclei.
The anisotropy is function of the anomalous fractal dimensions of the colliding objects.
The fractal dimensions characterize constituent fractal-like hadronic sub-structure
which seems to be universal property of hadronic matter
revealed at high energies.
Presented approach to the $z$ scaling shows that the observed
regularity has relevance to fundamental principles of physics at small scales.
More detailed study of the fractal aspects of $z$-scaling, both theoretical and experimental,
can give better understanding of the structure of hadrons and nuclei,
interaction of their constituents and particle formation
in the domain tested by large accelerators of hadrons and nuclei.

\vskip 0.5cm
{\section {Acknowledgments}}

The author thanks M.V. Tokarev and M. Vymazal on helpful discussions on this work.

\vskip 0.5cm
{\section {Appendix}}

In the theory of special relativity, the spatial symmetry lies at the
root of the standard pseudo-Euclidean metric.
Uni-directional breakdown of this symmetry compatible with the
relativistic methods was considered within the framework of Finsler
geometry \cite{Rund,Asanov1}.
Spatial anisotropy is expressed in terms of the Finslerian parameter $\vec{g}$
which is treated as an universal constant of pure geometrical origin.
It characterizes degree of Finslerian non-Riemannianity of space-time.
For small values of $\vec{g}$, the Finsler-relativistic metric function $F(\vec{g},R)$
and the associated Hamiltonian function $H(\vec{g},P)$ \cite{Asanov2}
can be approximated as follows
\begin{equation}
F(g,R) =
\vert T+g_{+}\vert \vec{R}\vert\vert^{(1+u)/2}
\vert T+g_{-}\vert \vec{R}\vert\vert^{(1-u)/2}\sim
(T^2-R^2-gT\vert\vec{R}\vert)^{1/2},
\label{eq:a1}
\end{equation}
\begin{equation}
H(g,P) =
\vert P_0-g_{+}\vert \vec{P}\vert\vert^{(1+u)/2}
\vert P_0-g_{-}\vert \vec{P}\vert\vert^{(1-u)/2}\sim
(P_0^2-P^2+gP_0\vert\vec{P}\vert)^{1/2}
\label{eq:a2}
\end{equation}
where we have used the notation
\begin{equation}
u = \frac{g/2}{\sqrt{1\!+\!(g/2)^2}} .
\label{eq:a3}
\end{equation}
Comparing the approximate expressions for the Finsler metric function (\ref{eq:a1})
and the associate Hamilton function (\ref{eq:a2})
with the invariants (\ref{eq:x53}) and (\ref{eq:x80}), one can judge to
the correspondence $g=2U$ between the Finslerian parameter $g$
and the quantity $U$ given by Eq. (\ref{eq:x15}). This is seen from the terms
proportional to $gT$ and $gP_0$, respectively.
While both parameters $g$ and $U$ characterize space-time isotropy violation,
there are substantial differences in the form of the metric invariants even for
their small values.
In the Finslerian case, the light front is approximated by a sphere with the radius
$\sqrt{1+(g/2)^2}$. The sphere is shifted in the direction of the isotropy
breakdown by a value $g/2$. Spherical form of the light front involves deformation of the
scales perpendicular to motion whenever $g\ne 0$ and thus results in violation of
the spatial isotropy. Similar concerns the Hamiltonian associate function (\ref{eq:a2})
and the accessible range of the corresponding ratio $\vec{P}/P_0$.
Unlike the Finslerian metric forms, the invariant (\ref{eq:x53}) is different.
The light front forms an ellipse with
one focus situated in the point $\vec{v}=0$. The elliptical form preserves the
scales perpendicular to motion even for $U\ne 0$. In the situation we consider, the
spatial isotropy is thus not violated.
Similar holds for the invariant (\ref{eq:x80})
and the accessible range of the corresponding parameter $\vec{P}/E$.

The important point in both approaches is that, in the regions of $g\ne 0$ $(U\ne 0)$,
the light velocity value should be  anisotropic in whatever inertial reference system.
Standard interpretation of the Michelson-Morley-type experiments
(including optical interferometer experiments \cite{Michelson,Kennedy,Ives} and modern
high-precision laser experiments \cite{Kirsher,Hils})
seems to be however negative with this respect.
The experiments steadily reproduce "no fringe shift" and, therefore, do not support
any deviation which would point to even tiny portion of the anisotropic spread
of light. In the Finslerian treatment, "null result" of these experiments was
interpreted \cite{Asanov2} as a "possible conspiracy of Nature" in the compensation
of two effects: the light-velocity anisotropy and the standard spatial length anisotropy.
Possibility of such compensation was shown to the first order of accuracy with
neglecting the second-order relativistic effects.
We show bellow that the
Michelson-Morley-type experiments alone do not imply absolute absence of anisotropy
in light propagation in our approach. The argumentation includes arbitrary
accuracy of relativistic effects.

Let us consider experiment with the interferometer having two perpendicular arms of the length
$d_I$ and $d_{II}$, respectively.
Suppose the light beam from a light source is divided into two rays, I and II,
traveling perpendicular to each other along the arms. The mirrors
placed on the ends of the spectrometer arms reflect the light
back to the telescope where the rays interfere with each other.
Assuming the apparatus is placed in a region where
the propagation of light is not isotropic one could expect
existence of a
phase difference $\Delta t$ between the rays I and II which is due
to the anisotropy. When the apparatus is rotated through an angle
of $90^0$, the orientation of the spectrometer arms is
interchanged and the phase difference becomes $-\Delta t$.
According to our standard intuition, such rotation of the apparatus should cause
a shift of the interference fringes between the two rays.
We show however, that this must not to be the case for any non-zero value of the
space-time anisotropy even up to the arbitrary order of experimental accuracy.
Suppose there exists a space-time anisotropy $\vec{u}$
induced by some reasons.
Let us assume that the anisotropy results in the metric changes
(\ref{eq:x52}) associated with deformation of the spherical light
front. In this case, the light front becomes an ellipsoid (\ref{eq:x71}) with
one focus in the point where the light was emitted (Fig.1).
The time $t_I$ and $t_{II}$ which the light rays take to travel
in spectrometer arms I and II can be expressed as follows
\begin{equation}
t_I   = d_I
\left(\frac{1}{v_1(\phi)} + \frac{1}{v_2(\phi)}\right), \ \ \ \ \
t_{II}= d_{II}
\left(\frac{1}{v_3(\phi)} + \frac{1}{v_4(\phi)}\right)
\label{eq:a4}
\end{equation}
where the angle $\phi$ describes orientation of the spectrometer with respect
to the space-time anisotropy $\vec{u}$. Because of the anisotropy, the
velocities of light propagation in different
directions $v_i(\phi)$ are not equal and depend on the orientation of the
spectrometer (Fig.1). On the other hand,  the spatial distances
(lengths of spectrometer arms) do not depend on the orientation
of the spectrometer in the metric (\ref{eq:x52}).
This follows from the known fact
that the spatial geometry is not simply given by the spatial part
$\eta_{ij}$ of the four dimensional metric
$\eta_{\mu\nu}(\vec{u})$.
The metric tensor $\eta^{\star}_{ij}$ which determines the
spatial geometry is given by \cite{Moller}
\begin{equation}
\eta^{\star}_{ij} = -\eta_{ij} + \eta^{\star}_i\eta^{\star}_j , \ \ \ \ \
\eta^{\star}_i = \frac{\eta_{i0}}{\sqrt{\eta_{00}}}.
\label{eq:a5}
\end{equation}
In the case of the four-dimensional metric (\ref{eq:x52}),
the spatial metric reads
\begin{equation}
\eta^{\star}_{ij} = \delta_{ij}.
\label{eq:a6}
\end{equation}
Therefore, lengths of the spectrometer arms (the distances $d_I$ and $d_{II}$)
are invariant under space rotations and thus do not depend on the angle $\phi$.
Note that the same holds for the metrics (\ref{eq:x81}),
$\bar{\eta}^{\star}_{ij} = \delta_{ij}$.

We exploit now the following geometrical property of the
velocity ellipsoid (\ref{eq:x71}).
While the sections $v_i(\phi)$ connecting any point of the ellipsoid with its focus
depend on their orientation $\phi$, the combinations
\begin{equation}
\frac{1}{v_1(\phi)} + \frac{1}{v_2(\phi)}
= \frac{2a}{b^2} , \ \ \ \ \ \
\frac{1}{v_3(\phi)} + \frac{1}{v_4(\phi)}
= \frac{2a}{b^2}
\label{eq:a7}
\end{equation}
are rotationally invariant i.e. do not depend on the angle
$\phi$. Here
\begin{equation}
a=\gamma^2, \ \ \ \ \
b=\gamma
\label{eq:a8}
\end{equation}
are major and minor semi-axis of the ellipsoid (\ref{eq:x71}), respectively.
After inserting expressions (\ref{eq:a7}) into
Eq. (\ref{eq:a4}), we get
\begin{equation}
t_I   = 2d_I, \ \ \ \ \ \
t_{II}= 2d_{II}.
\label{eq:a9}
\end{equation}
These relations connect time the light rays take to travel in the
spectrometer arms with the lengths $d_I$ and $d_{II}$ to the point they interfere.
Both expressions are rotationally invariant.
Therefore, rotation of the spectrometer apparatus can not
cause any shift of the interference fringes even for $\vec{u}\ne 0$.

In order to show that the invariance (\ref{eq:a7}) is not accidental we will
discuss more complicated case.
Let us consider a tree mirrors set-up which
reflect the rays of light along the sides of a triangle $ABC$. For definiteness consider the
triangle depicted by the full lines in Fig.2a. Suppose a light signal
is emitted in the point $A$ and then travels along
the path $d_1$, $d_2$, and $d_3$.  The corresponding time
interval
\begin{equation}
t_{ABC}   =
\frac{d_1}{v_1(\phi)} + \frac{d_2}{v_2(\phi)}
+ \frac{d_3}{v_3(\phi)}
\label{eq:a10}
\end{equation}
is function of the velocities $v_1(\phi)$, $v_2(\phi)$, and $v_3(\phi)$.
The velocities are shown on the velocity diagram in Fig.2b.
They depend on the orientation $\phi$ of the triangle $ABC$
relative to the space-time anisotropy $\vec{u}$.
We show that if this experimental set-up rotates, the time
$t_{ABC}$ remains invariant, though
values of the velocities $v_i(\phi)$ change during such rotations.
The $\phi$ invariance of the expression (\ref{eq:a10}) follows from the specific
geometrical property of any rotational ellipsoid which we outline below.
Let us denote the internal angles of the triangle $ABC$ as
$\alpha_1$,  $\alpha_2$, and $\alpha_3$ (Fig.2a). They comply the
elementary geometrical property of the constant ratio
\begin{equation}
\frac{d_1}{\sin\alpha_1}
=\frac{d_2}{\sin\alpha_2}
=\frac{d_3}{\sin\alpha_3} \equiv d_{ABC}
\label{eq:a11}
\end{equation}
which we denote as $d_{ABC}$.
The angles among the corresponding velocities $v_1(\phi)$, $v_2(\phi)$, and $v_3(\phi)$
are shown in Fig.2b and are indicated by $\beta_1$, $\beta_2$, and $\beta_3$, respectively.
In the considered mirror setup, the angles are fixed by the relations
\begin{equation}
\beta_i = \pi - \alpha_i,\ \ \ \ \ \ \ i = 1,2,3.
\label{eq:a12}
\end{equation}
Because of spatial rotational invariance (\ref{eq:a6}), the angles $\alpha_i$,
$\beta_i$, as well as the distances $d_i$ do not depend on the rotation of the apparatus
as the whole and thus do not depend on the angle $\phi$.
Therefore Eq. (\ref{eq:a10}) takes the form
\begin{equation}
t_{ABC}   = d_{ABC}\left(
\frac{\sin\beta_1}{v_1(\phi)} + \frac{\sin\beta_2}{v_2(\phi)}
+ \frac{\sin\beta_3}{v_3(\phi)}\right).
\label{eq:a13}
\end{equation}
Let us now exploit the following geometrical property valid for any rotational ellipsoid.
Consider the ellipse which forms intersection of the ellipsoid with a plane passing though
its focus. This focus is common focus for the ellipse and the ellipsoid as well.
Moreover, for any orientation of this plane,
\begin{equation}
\frac{A}{B^2}=\frac{a}{b^2}
\label{eq:a14}
\end{equation}
where $A$ and $B$, or $a$ and $b$ are major and minor semi-axes of the ellipse
or the ellipsoid, respectively.
Here we have in mind the velocity ellipsoid (\ref{eq:x71})
in space-time with the anisotropy $\vec{u}$ and the plane determined by the orientations of
the ray velocities passing through the arms of the triangle $ABC$.
The ray velocities mark out three different points on such ellipse.
Let us denote the sections connecting the focus of the ellipse with these
points by $v_1(\phi)$, $v_2(\phi)$, and $v_3(\phi)$, respectively (Fig.2b.).
One can convince itself that, while the magnitudes of
the ray velocities $v_i(\phi)$ are functions of the angle $\phi$, the combination
\begin{equation}
\frac{\sin\beta_1}{v_1(\phi)} + \frac{\sin\beta_2}{v_2(\phi)}
+ \frac{\sin\beta_3}{v_3(\phi)}
= \frac{a}{b^2}\left(\sin\beta_1+\sin\beta_2+\sin\beta_3\right)
\label{eq:a15}
\end{equation}
does not depend on  $\phi$.
The symbols $a$ and $b$ are given by (\ref{eq:a8}) and denote the
major and minor semi-axis of the ellipsoid (\ref{eq:x71}), respectively.
Searching for the above remarkable geometrical property of the rotational ellipsoids was
inspired by pure physical reasons and shows how physics and geometry are tightly
interconnected. The relation (\ref{eq:a15}) represents continuous  generalization
of the expressions (\ref{eq:a7}).
Really, if we identify the point $A$ with the point $B$ of the triangle $ABC$,
it degenerates into the abscissa $AB$. In this case $d_3=0$, $\alpha_3=0$,
$\alpha_1=\alpha_2=\pi/2$ and the relation (\ref{eq:a15}) becomes
identical with Eq. (\ref{eq:a7}).
Now it remains to exploit Eqs. (\ref{eq:a8}),  (\ref{eq:a11}), (\ref{eq:a13}),
(\ref{eq:a15}), and one gets the expression
\begin{equation}
t_{ABC}   = d_1 + d_2 + d_3
\label{eq:a16}
\end{equation}
which does not depend on the orientation $\phi$.
Therefore, time $t_{ABC}$ the light rays take to travel along the triangle $ABC$
does not depend on its orientation with respect to the space-time anisotropy
$\vec{u}$.
As a consequence, arbitrary rotation of a three mirror set-up will not
cause any shift of the interference fringes of light even for $\vec{u}\ne 0$.

Let us now imagine the light signal traveling along the
triangles $ABC$ and $ACD$ depicted in Fig.2a in the following order.
The signal is emitted in the point $A$ and travels along the lines
$d_1$, $d_2$, and $d_3$. The signal is partially reflected back in the
point $A$ and then travels the distances $d_4$, $d_5$, and $d_6$.
As follows from the above considerations applied to both triangles $ABC$ and $ACD$
separately, the light takes to travel the whole path during time
\begin{equation}
t_{ABC} + t_{ACD}  = d_1 + d_2 + d_3 + d_4 + d_5 + d_6 =
t_{out}+t_{int}
\label{eq:a17}
\end{equation}
which does not depend on the particular choice of the angle $\phi$.
Here we have denoted by $t_{int}$ time the light ray travels
along the internal line $CA$ to and fro.
According to Eq. (\ref{eq:a9}), $t_{int}$  depends on the distance
$d_{CA}$ through the rotational invariant relation $t_{int}=d_3+d_4 =2d_{CA}$.
Consequently, the expression
\begin{equation}
t_{out} = d_1 + d_2 + d_5 + d_6
\label{eq:a18}
\end{equation}
possess the rotational symmetry and does not depend on the space-time
anisotropy $\vec{u}$, as well.
It is possible to think of various trajectories from the
point $A$ to the point $B$ corresponding to various experimental arrangements of
interference experiments.
The "null result" of the interference fringe shift with rotations can be shown
for such trajectories similarly.

This appendix should be understand so that we do not advocate
the anisotropic  spread of light in general. We point here only to the
fact that Michelson-Morley-type interferometer experiments do not contradict
to particular situations in which the anisotropy of light propagation in space-time
could not be a'priori excluded.
This concerns not only small scale structures in particle physics but also
space-time features  at cosmological distances.
There exists new theoretical studies \cite{Bereca} suggesting
eccentric expansion of the universe resulting from its arrangements at large scales.
These investigations point to possible non-sphericity
of the universe which at some $t$ could evolve into an ellipsoid.

\newpage
\begin{center}

\begin{figure}
\epsfig{file=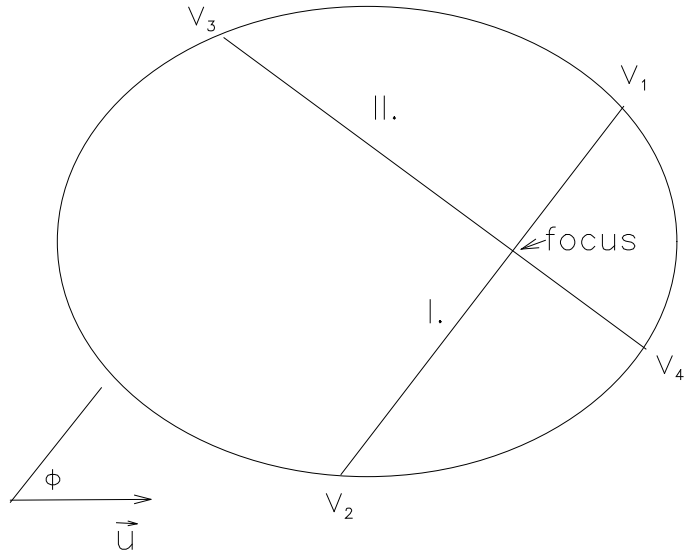, width=90mm}
\vspace*{5mm}
\caption{The velocity diagram in space-time with the structural anisotropy
$\vec{u}$. The lines I. and II. correspond to the orientation
of the spectrometer arms in the Michelson's experiment.
}
\end{figure}

\newpage
\begin{figure}
\epsfig{file=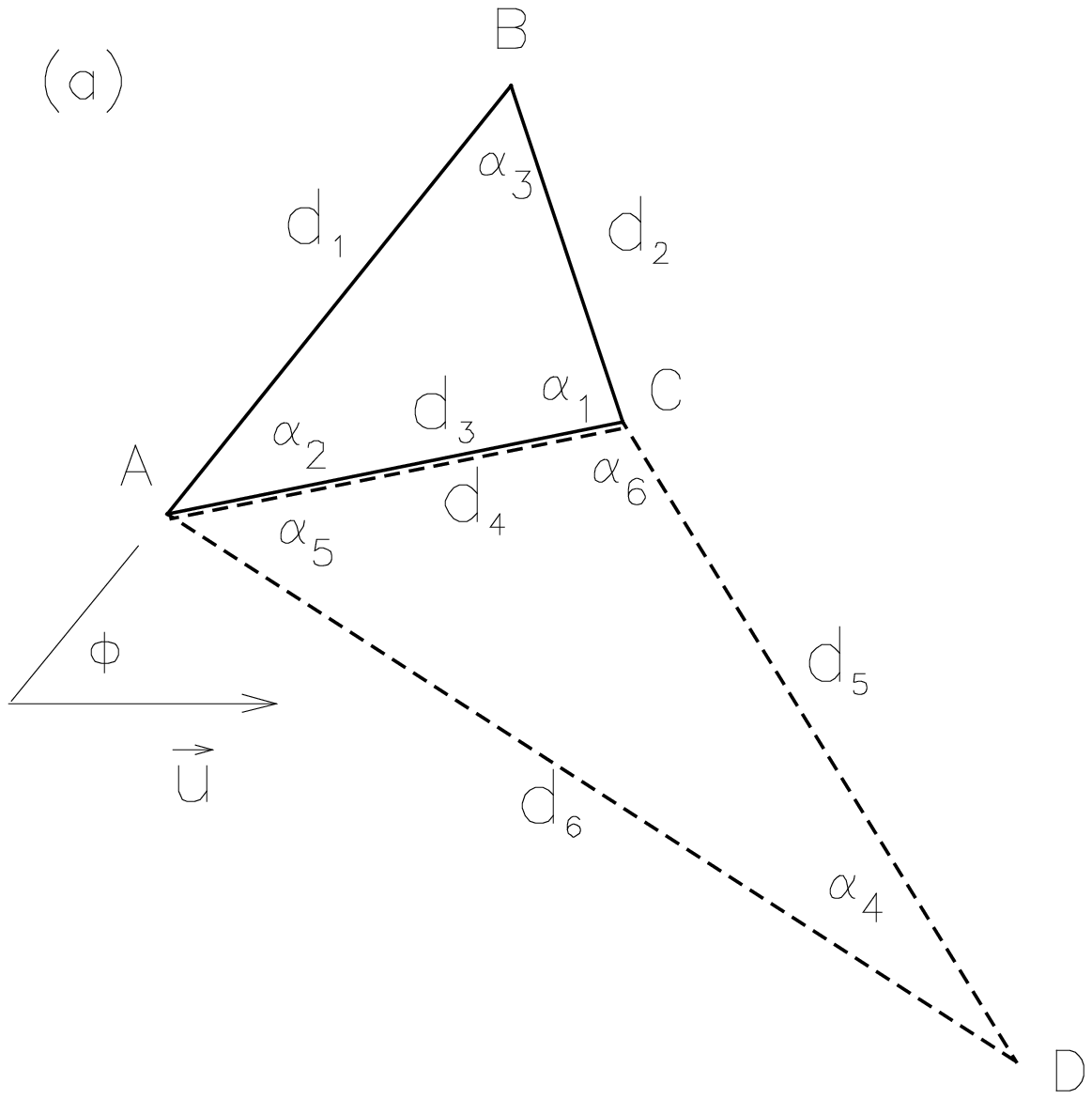, width=90mm}
\end{figure}
\vspace*{10mm}
\begin{figure}
\epsfig{file=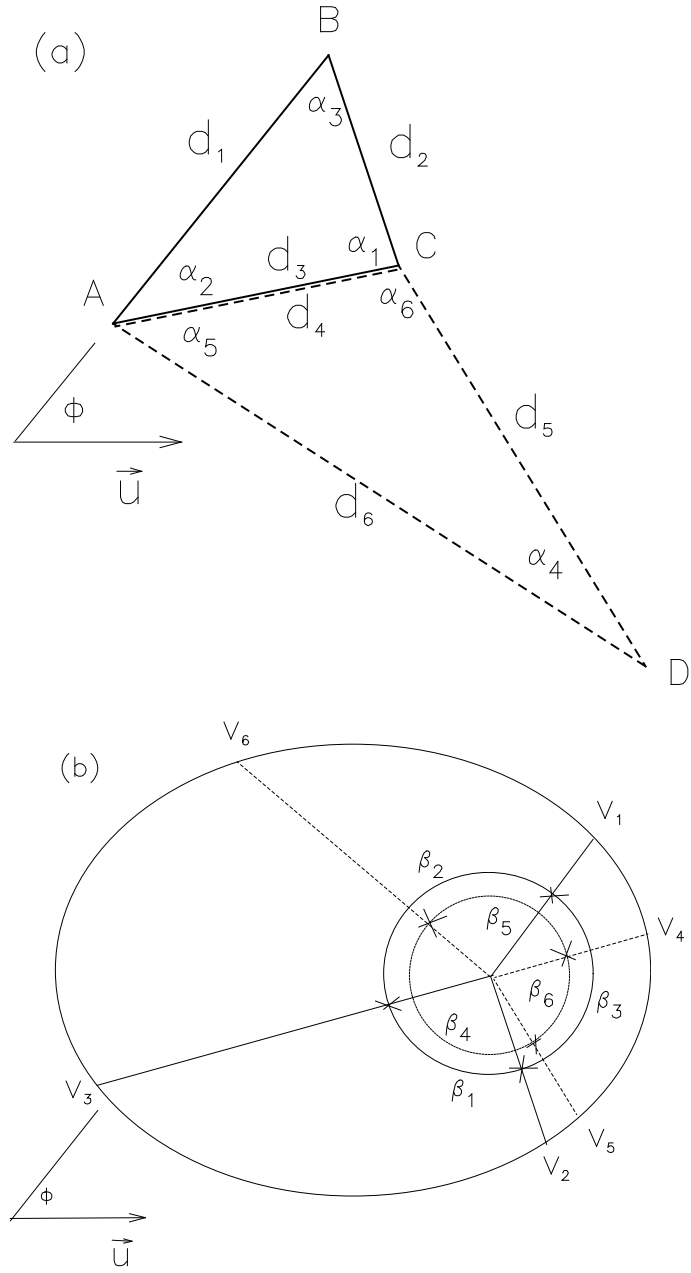, width=90mm}
\vspace*{10mm}
\caption{
(a) The space diagram of a multi-mirror setup. The mirrors
are considered in the points $A$, $B$, $C$, and $D$ reflecting
the light signal along the sketched lines.
(b) The velocity diagram corresponding to the mirrors
    setup shown in Fig.2a.
}
\end{figure}
\end{center}

\end{document}